\documentclass[prd,aps,twocolumn,nofootinbib]{revtex4}
\usepackage{graphicx}
\usepackage{latexsym,amssymb,amsmath,amsfonts,amssymb,txfonts,pxfonts,wasysym,float}

\usepackage{wasysym}
\usepackage{bm}
\usepackage[usenames,dvipsnames]{xcolor}
\definecolor{orange}{cmyk}{0,0.5,1,0}
\definecolor{rossoCP3}{cmyk}{0,.88,.77,.40}
\definecolor{graa}{rgb}{0.8,0.8,0.8}
\definecolor{blaa}{rgb}{0.2,0.2,0.6}

\def\gr{$\gamma$-ray}

\begin{document}
\title{ Sensitivity of a proposed space-based Cherenkov astrophysical-neutrino
   telescope (CHANT) }
\author{Andrii Neronov}
\affiliation{Astronomy Department, University of Geneva, Ch. d'Ecogia 16, 1290, Versoix, Switzerland}
\author{Dmitri V. Semikoz}
\affiliation{APC, Universite Paris Diderot, CNRS/IN2P3, CEA/IRFU,\\
Observatoire de Paris, Sorbonne Paris Cite, 119 75205 Paris, France and\\
National Research Nuclear University MEPHI (Moscow Engineering Physics Institute), Kashirskoe highway 31, Moscow, 115409, Russia}
\author{Luis A. Anchordoqui}
\affiliation{Department of Physics and Astronomy, Lehman College, City University of
  New York, NY 10468, USA\\
Department of Physics, Graduate Center, City University
  of New York, 365 Fifth Avenue, NY 10016, USA\\
Department of Astrophysics, American Museum of Natural History, Central Park West  79 St., NY 10024, USA}
\author{James H. Adams}
\affiliation{University of Alabama in Huntsville
301 Sparkman Drive, Huntsville, AL 35899, USA}
\author{Angela V. Olinto}
\affiliation{Department of Astronomy \& Astrophysics, Kavli Institute for Cosmological Physics, The University of Chicago, Chicago, IL 60637, USA}

\begin{abstract} 
\noindent  Neutrinos with energies in the PeV to EeV range produce upgoing extensive air showers when they interact underground close enough to the surface of the Earth. We study the possibility for detection of such showers with a system of very wide field-of-view imaging atmospheric Cherenkov telescopes, named CHANT for CHerenkov from Astrophysical Neutrinos Telescope,  pointing down to a strip below the Earth's horizon from space. We find that CHANT provides sufficient sensitivity for the study of the astrophysical neutrino flux in a wide energy range, from 10~PeV to 10~EeV.  A space-based CHANT system can discover and  study in detail the cosmogenic neutrino flux originating from interactions of ultra-high-energy cosmic rays in the intergalactic medium.  \end{abstract} \maketitle

\section{Introduction}

The IceCube neutrino detector has recently marked the start of neutrino astronomy by reporting the discovery of an astrophysical neutrino signal. The  spectrum  of the signal, detected in the all-flavour high-energy starting events (HESE) with neutrino vertex contained in the detector volume, is  consistent with a power-law 
\begin{equation}
\frac{dN_\nu}{dE}=A\times 10^{-18}\left[\frac{E}{10^{14}\mbox{ eV}}\right]^{-p}\frac{1}{\mbox{ GeV cm}^2\mbox{ s sr}} \,,
\end{equation}
within a broad energy range, 30~TeV$<E<2$~PeV, and without signature of a high-energy cut-off ~\cite{IceCube_combined_2015}.  The best-fit power law corresponds to 
a normalisation  $A=6.7^{+1.1}_{-1.2}$ and spectral index $p = 2.5\pm 0.09$. A somewhat harder slope $p = 2.2\pm 0.2$ is found in the muon neutrino signal for upgoing muon neutrinos detected above $\sim 100$~TeV~\cite{IceCube_muonnu_2015,IceCube_ICRC,icecube_6yr_muon}. The highest energy neutrino event observed thus far is an upward-moving muon track with zenith angle of $101^\circ$ with energy deposited  $(2.6 \pm 0.3) \times 10^6~{\rm  GeV}$~\cite{Aartsen:2016ngq}. It has been speculated that this event may have been triggered by a $\nu_\tau$, with an energy beyond 10~PeV and possibly up to 100~PeV~\cite{multi-PeV1}.

Low statistics of the signal prevents identification of its origin. It is possible that at least part of the flux originates in the Milky Way, as indicated by the $3\sigma$ evidence for anisotropy along  the direction of the Galactic plane~\cite{neronov} and by the consistency of the neutrino spectrum with the spectrum of diffuse Galactic \gr\ emission \cite{neronov1}. It has also been suggested that some of the events may have originated from the Galactic center region~\cite{Razzaque:2013uoa,Bai:2014kba,Anchordoqui:2016dcp}. However, shower events have poor angular resolutions to reveal their origin and so 4~yr data set is also consistent with isotropy~\cite{IceCube_ICRC}. A complementary indication for the  existence of a Galactic component could also be derived from a tension between the measurement of the neutrino flux in the HESE~\citep{IceCube_combined_2015} and muon track neutrino samples   \cite{IceCube_muonnu_2015,IceCube_ICRC}. These two event types sample different parts of the sky,  with the HESE events having more exposure in the direction of the inner Galaxy and the muon neutrino flux sampling a low declination strip~\cite{neronov_muon}. Finally, Waxman-Bahcall energetics does not exclude a Galactic origin for a significant component of the IceCube flux~\cite{Anchordoqui:2013qsi}.

Independently of the presence or absence of the Galactic component of the astrophysical neutrino signal, a significant contribution to the flux could come from a population of extragalactic cosmic ray  sources \cite{ahlers_review,giacinti15} with active galactic nuclei (AGN) \cite{Stecker91,mannheim92,neronov_semikoz02,tchernin13} and star-forming and/or starburst galaxies \citep{loeb06,murase,tamborra,giacinti15,no_starforming} being the most discussed candidate source classes. For a review, see e.g.~\citep{anchordoqui13}.

Still another type of the astrophysical neutrino signal is generically expected to be produced by  the interactions of ultra-high-energy cosmic rays (UHECR) with low energy photon backgrounds: the cosmic microwave background (CMB) and the extragalactic background light (EBL) \citep{berezinsky,stecker73_cosmogenic,berezinsky1,engel01,semikoz_sigl}. The flux level and spectral shape of this ``cosmogenic'' neutrino flux is determined by  the elemental composition of the UHECR flux (proton or heavy nuclei dominated) and by the cosmological evolution of the UHECR source population \citep{allard06,kotera10,ahlers10,aloisio15}.    

A conventional approach for the detection of neutrinos with energies in the PeV range is based on detectors which use large volumes of ice (IceCube\footnote{https://icecube.wisc.edu}) or water (ANTARES\footnote{http://antares.in2p3.fr}) as the detector medium and sample Cherenkov light from tracks of muons produced by muon neutrinos, or from electron and tau lepton induced showers initiated by the charged current interactions of electron and tau neutrinos. Upgrades of existing detectors, such as IceCube  Generation 2~\citep{IceCube_gen2}, and deployment of new detectors, such as  KM3NeT~\cite{km3net} and GVD~\cite{GVD}, are expected to bring an improvement of sensitivity and increase of statistics of the astrophysical neutrino signal. Another technique which potentially provides higher sensitivity in the PeV-EeV energy range is that of detection of radio signal from the neutrino-induced  particle showers in the detector medium. This technique is used by RICE \cite{RICE} and is planned for the ARA ~\cite{ARA} and   ARIANNA~\cite{ARIANNA} arrays.

An alternative technique is based on the observation of upgoing air showers (UAS) produced by the leptons originating from neutrino interactions below the Earth surface. It was previously introduced as a possible technique for detection of ultra-high-energy tau neutrinos \citep{fargion02,feng02,bertou,kusenko,aramo,auger15}.    The neutrino induced UAS could be detected using a variety of detection methods: surface particle detector arrays and air fluorescence telescopes, like in the Pierre Auger Observatory~\cite{auger15} and the Telescope Array\footnote{http://www.telescopearray.org}, or radio detectors like ANITA~\cite{ANITA} and  the planned   EVA~\cite{EVA} and  GRAND array \cite{grand}. Still another possibility is to use the technique of  Imaging Atmospheric Cherenkov Telescopes (IACT) which is widely used in comtemporary \gr\ astronomy. An IACT system capable to detect the neutrino-induced UAS should overlook the ground, e.g., the side of a mountain \cite{ashra,gora15,gora16}. 

In this paper we show that observations of  $\nu_\tau$ induced UAS with an IACT system, named CHANT for for CHerenkov from Astrophysical Neutrinos Telescope, can be used  for the study of the astrophysical neutrino flux at energies as low as 10~PeV  and as high as tens of EeVs. The required sensitivity  for detection of the astrophysical neutrino flux can be reached by lifting a telescope to high altitude, the top of a high mountain or on an ultra-long duration balloon flight, to monitor a large surface on the ground. The optimum possibility is to deploy the telescope in space to maximize  the monitored surface area. Below we consider the details of the space-based CHANT system and discuss the high-altitude and balloon alternatives at the end. CHANT will also have sufficient sensitivity for the discovery and study of the challenging cosmogenic neutrino flux, even if the UHECR flux is dominated by heavy nuclei.

{Throughout we assume that tau neutrinos constitute 1/3 or the overall neutrino flux from astronomical sources.  This is
expected under the conventional assumptions that neutrino production proceeds via the decay chain
\begin{eqnarray}\label{eqn:chain}
\pi^+ \to & \mu^+ & \nu_\mu ~~~~~~~~~~~~~~~~{\rm (and \, the \,
  conjugate \, process)} \,, \\
                & \reflectbox{\rotatebox[origin=c]{180}{$\Rsh$}} &
                 e^+ \, \bar \nu_\mu \, \nu_e \nonumber \end{eqnarray} 
and that there is maximal mixing between muon and tau neutrinos. The decay chain may be complete in the sense that both decays indicated in Eq.~(\ref{eqn:chain}) occur without significant change in the $\mu$ energy, or it may be incomplete, in which case the $\mu$ suffers possibly catastrophic energy loss before decay.  Note that the minimum flux of tau neutrinos corresponds to the complete decay chain hypothesis. This is because taking the global best-fit mixing parameters~\cite{Gonzalez-Garcia:2014bfa}, a flavor ratio at Earth of $(f_{\nu_e}:f_{\nu_\mu}:f_{\nu_\tau}) = (0.93:1.05: 1.02)_\oplus \approx (1:1:1)_\oplus$ is expected for a $(1:2:0)_S$ source composition. For the incomplete decay chain, with flavor ratios at the source $(0:1:0)_S$, the composition at Earth becomes $(0.6:1.3:1.1)_\oplus$. Antineutrinos can also be produced via neutron $\beta$-decay, yielding $(1:0:0)_S$ and after oscillations {\it en route} to Earth $(1.6:0.6:0.8)_\oplus$~\cite{Anchordoqui:2003vc}. However, existing data  disfavor a $(1:0:0)_S$ flavor ratio (with best fit $(0:0.2:0.8)_\oplus$)~\cite{Aartsen:2015ivb}, and so our original assumption that $\nu_\tau$ constitute a third of the total neutrino flux is conservative.}

\section{Neutrino Detection  with imaging atmospheric Cherenkov telescopes}

\subsection{$\bm{\nu_\tau}$ and $\bm{\tau}$ propagation ranges}

Tau neutrino interacts inside the Earth on a distance scale 
\begin{eqnarray}
  \lambda_\nu \! \! & = & \! \! \frac{m_p}{\sigma_{\nu N}\rho} \nonumber \\
  & \simeq &\!\! 1.6\times 10^8\left[\frac{E}{10^{17}\mbox{ eV}}\right]^{-0.3}\left[\frac{\rho}{2.65\mbox{ g/cm}^3}\right]^{-1} \! \! \! {\rm cm}, \end{eqnarray} 
where $\sigma_{\nu N}\simeq 4\times 10^{-33}\left(E/10^{17}\mbox{ eV}\right)^{0.3}$~cm$^2$ is the interaction cross-section \cite{cross_section_loss} and $\rho$ is the density of the Earth medium (in what follows we adopt $\rho=1$~g/cm$^3$ for water and $\rho=2.7$~g/cm$^3$ for rock). A point worth noting at this juncture is that for $E \alt 1~{\rm EeV}$, perturbative
  QCD provides a robust framework to calculate the neutrino-nucleon
  cross section~\cite{Gandhi:1995tf,CooperSarkar:2007cv,Connolly:2011vc,Arguelles:2015wba}.  It is only
  when the fractional momenta $x$ carried by the nucleon constituents become
  vanishingly small that the structure functions develop a $\ln(1/x)$
  divergent behavior, which in turn results in a violation of unitarity
  bounds. Consequently, perturbative QCD predictions are expected to
  break down solely when the nucleon has an increasing number of partons with
  small $x$.  For the center of mass energies
  relevant to our study, however, the neutrino-nucleon cross section can
  be calculated perturbatively with an accuracy of better than 10\% when
  constrained by measured HERA structure
  functions, see e.g. Fig~13 in Ref.~\cite{CooperSarkar:2011pa}.

The neutrino interactions could result in the production of an UAS if the interaction vertex is close enough to the Earth surface, within the propagation range of the tau lepton. Tau leptons with energies below $E\lesssim 10^{17}$~eV propagate over the decay distance range \begin{equation} \label{eq:lambda} \lambda_{\tau}=5\times 10^{5}\left[\frac{E}{10^{17}\mbox{ eV}}\right]\mbox{ cm} \, .  \end{equation}

At higher energies, energy loss processes become important. The most important ones being the photonuclear scattering and $e^+e^-$ pair production. Several groups of authors starting from Ref.~\cite{Fargion:2000iz} calculated those processes both, analytically assuming a continue loss approximation, and  numerically using the stochastic approach~\cite{Aramo:2004pr,Dutta:2005yt,Bigas:2008ff}.  Here we adopt the results of the stochastic calculation of Ref.~\cite{Dutta:2005yt}, which show  that the characteristic distance on which a $\tau$ lepton loses half of its energy is  
\begin{equation}
\label{eq:ltau}
l_\tau \sim 3\times 10^5\chi(\rho ,E)\mbox{ cm} \,,
\end{equation}
where $\chi\sim 1$ is a numerical factor which depends on the elemental composition of the medium and theoretical uncertainties from the calculation of the interaction cross-sections~\cite{cross_section_loss}. For rock, $\chi\simeq 1$, and is twice as large for water. 
The energy loss distance becomes much shorter than the decay distance at the energies above $\sim 10^{17}$~eV. Taus produced within the $\min(\lambda_\tau,l_\tau)$ distance below the surface  emerge in the atmosphere and decay producing an extensive atmospheric shower (EAS) of either hadronic (in $\simeq 65$\% cases) or electromagnetic (with $\sim 18\%$ probability) nature. In about 17\% of the cases the tau lepton decays with production of a muon  without an associated EAS~\citep{pdg,fargion02}.

\subsection{Telescope setup}

\begin{figure} \includegraphics[width=\linewidth]{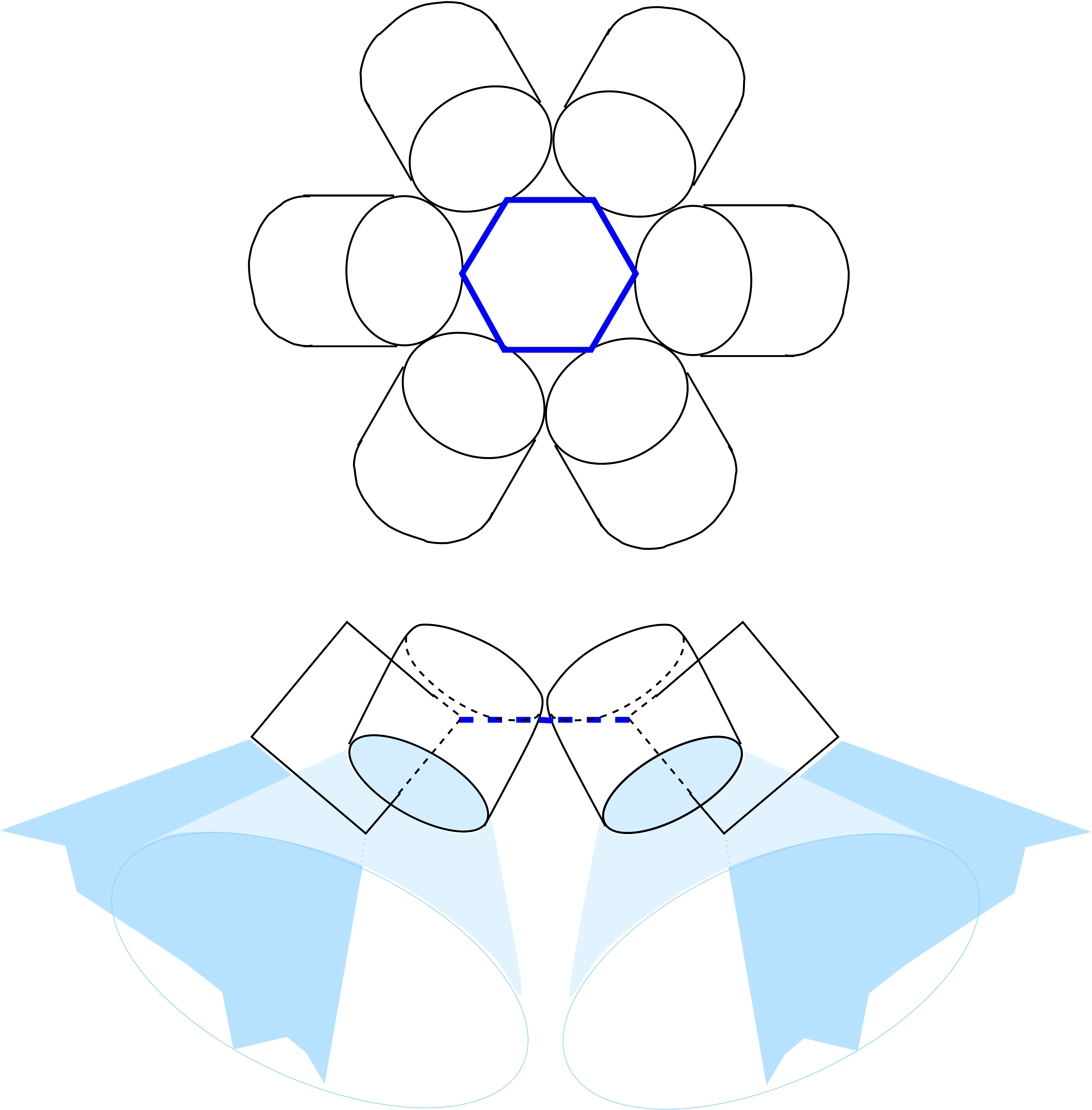} \includegraphics[width=\linewidth]{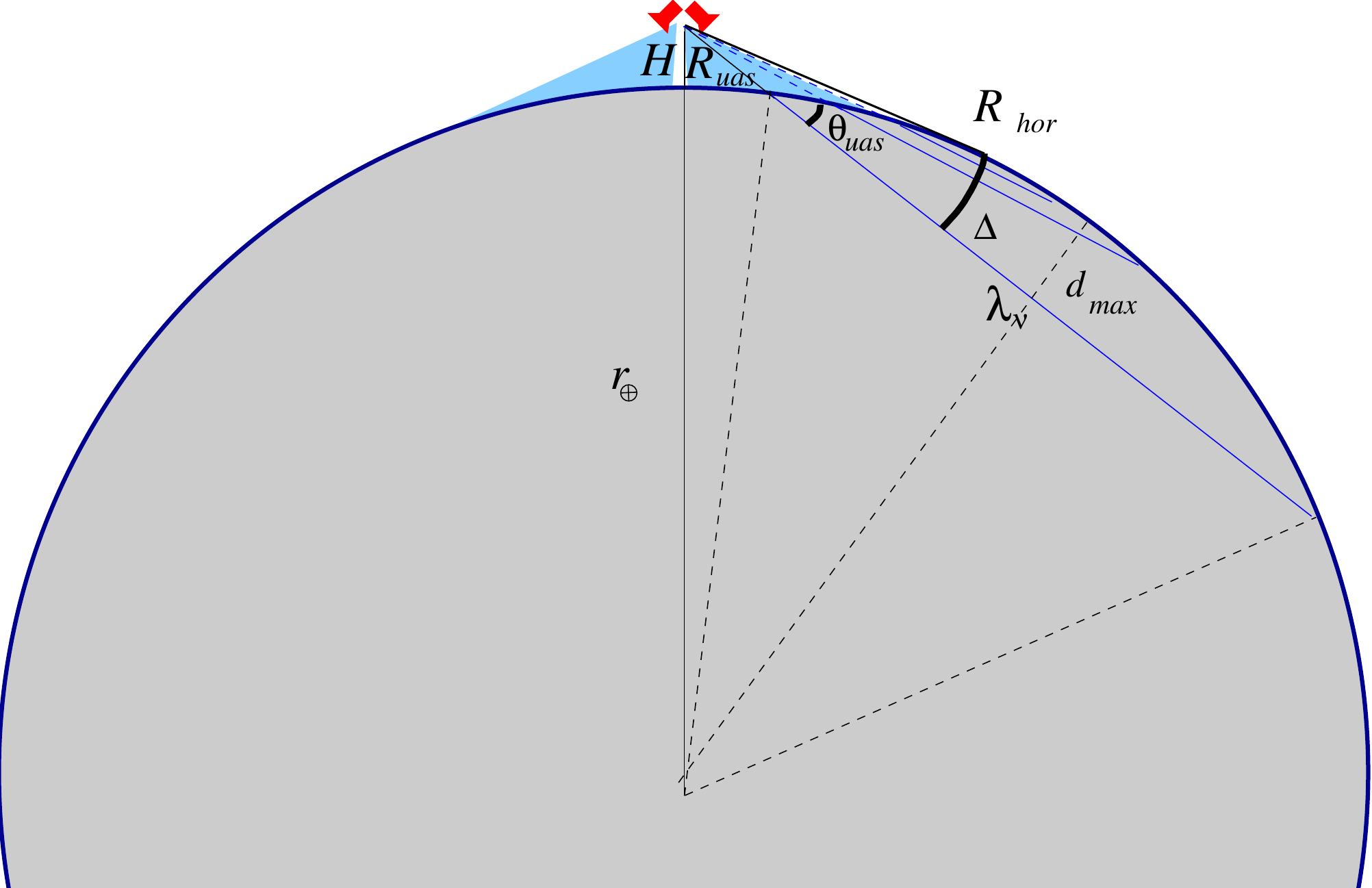} \caption{The bottom panel illustrates the neutrino detection principle with a space or balloon borne CHANT system. The top and middle panels exhibit  top and side views of possible arrangement for the telescope modules providing $360^\circ$ overview of the strip below the Earth limb. Each CHANT module is a refractor telescope with $60^\circ$ wide FoV.}
\label{fig:tilt_mode}
\end{figure}

The mean free path of neutrinos with energies above PeV
 is shorter than the mean Earth radius $r_\oplus=6371$~km. Thus, the neutrino induced UAS would typically  have an elevation angle 
\begin{equation}
\theta< \theta_{\rm uas}=\arcsin\left(\frac{\lambda_\nu}{2r_\oplus}\right)\simeq7^\circ 
\left[\frac{E}{10^{17}\mbox{ eV}}\right]^{-0.3} \, .
\end{equation} 
The UAS emerging at largest elevation angles are produced by neutrinos passing unobscured at the maximal depth 
\begin{eqnarray}
d_{\rm max}\le \frac{\lambda_\nu^2}{8r_\oplus}\simeq 
50
\left[\frac{E}{10^{17}\mbox{ eV}}\right]^{-0.6}\mbox{ km} \, .
\end{eqnarray}
The angular width of a strip containing the detectable neutrino flux is about
\begin{eqnarray}
\Delta & \simeq & \arcsin\left(\frac{d_{\rm max}}{R_{\rm hor}}\right) \nonumber \\
&\sim & 2^\circ\left[\frac{E}{10^{17}\mbox{ eV}}\right]^{-0.6}\left[\frac{H}{300\mbox{ km}}\right]^{-0.5} \, ,
\label{eq:Delta}
\end{eqnarray}
where \begin{equation}
\label{eq:r_hor}
R_{\rm hor}\simeq \sqrt{2r_\oplus H}\simeq 2\times 10^3\left[\frac{H}{300\mbox{ km}}\right]^{0.5}\mbox{ km}
\end{equation}
is the distance to the horizon from a telescope situated at an altitude $H$. Note that the numerical estimate in Eq. (\ref{eq:Delta}) is not valid in the energy range below  $10^{17}$~eV because $\Delta$ becomes macroscopically large. 

The low elevation angle UAS could be detected by an IACT pointed at the strip of angular width $\Delta$ just  below the Earth horizon, as it is shown in the bottom panel of Fig. \ref{fig:tilt_mode}. Obviously, the neutrino signal is maximised if a system of several wide Field-of-View (FoV) telescopes monitors the full $360^\circ$ strip below the horizon. 
Such a system could be realised as shown in the top and middle  panels of Fig. \ref{fig:tilt_mode}.  
These two panels show top and side views of a system consisting of six EUSO-type \cite{jemeuso} refractor telescopes with the FoV $60^\circ$ wide. Each telescope is tilted with respect to the nadir direction, to have a $\gtrsim 60^\circ$ wide part of the horizon strip in its FoV. Observations of neutrinos  require the FoV of each telescope module to be $60^\circ\times \Delta$, i.e., only a small part of the full available FoV $60^\circ$ in diameter is used for detection of the highest energy neutrinos. 

A similar arrangement could be realised also with reflector type telescopes, conventionally used in the Cherenkov astronomy. However, this would possibly require a larger number of telescope modules, because the FoVs of reflector IACTs are typically much more narrow. An advantage of the reflector system is its larger optical throughput (close to $1$, compared to $\simeq 0.5$ of the EUSO type optical system \cite{jemeuso}). Another advantage is technological: large optical reflectors are easier to produce, deploy, and  operate than large multi-lens refractor systems.  Still another option that could be used is a system employing modules with catadioptric {Schmidt camera} optical system like that of the ASHRA telescope \cite{ashra} or the optical system of OWL telescopes \cite{OWL}. This system potentially provides wide FoV and larger optical throughput combined with wider FoV. 

Overall, the full focal surface instrumentation of the six modules of the CHANT system would be equivalent in the number of pixels, complexity,  weight, and  cost to that of the EUSO telescope.  This assumes each of the six modules uses only a narrow strip of the area, 1/6 of the EUSO focal surface. To the contrary, a more complex optical system of CHANT (six modules instead of one, compared to EUSO) would lead to an increase of the mission cost and mass. In EUSO, the optics sub-system contributes about 19\% of the instrument cost and mass budget\footnote{JEM-EUSO Phase A study report http://euso-balloon.lal.in2p3.fr/IMG/pdf/PurpleBook\_2010\_v-5\_MCM\_lightreso.pdf}. Taking this estimate as an indication for the CHANT system adopting EUSO optical design, one could find that the optical system of the six modules of CHANT would contribute about half of the instrument mass and cost.  

Below we consider a reference design with parameters summarized in Table \ref{tab:params}. 

\begin{table}
\label{tab:params}
\begin{tabular}{|l|c|}
\hline
Parameter & Range of values\\
\hline
Number of telescopes & 6\\
Telescope aperture $D_{tel}$ & 2~m -- 4~m\\
Telescope Field-of-View & $60^\circ\times 15^\circ$\\
Pixel size & $0.1^\circ$\\
Number of pixels (per telescope) & $9\times 10^4$\\
Optical efficiency $\epsilon$ & 0.1\\
Observation wavelength & {500-600} nm\\ 
Observation altitude & 3~km -- 300 km\\
\hline
\end{tabular}
\caption{Basic parameters of the CHANT telescope system.}
\end{table} 

\subsection{Use of the atmosphere as a giant particle detector}

Similarly to other Cherenkov telescopes, CHANT will use the atmosphere as a particle detector. Proper interpretation of the EAS data would require control over the state of this detector, in particular of the optical properties of the atmosphere. These optical properties are affected by the presence of clouds and aerosols. Characterisation of the properties of the clouds and aerosols will require a dedicated Atmospheric Monitoring System, similar to that of the EUSO telescope \cite{euso_ams}. This includes on-board instruments like LIght Detection And Ranging (LIDAR) device and an infrared camera complemented by the information provided by the Global Atmospheric monitoring data collected for meteorology and atmospheric research. 

Contrary to EUSO which observes EAS events from the distance $\sim H$ in nadir direction, CHANT would detect strongly inclined upgoing EAS which occur at the distances about $R_{\rm hor}$. Larger distance and higher atmospheric column density in the direction of such events would require a more powerful laser for the LIDAR operating in the visible, rather than UV band. For example, a balloon-borne CHANT system would detect EAS at the distance $R_{\rm hor}\sim 600$~km (see Eq.~(\ref{eq:r_hor})) comparable to that of the observation distance of EUSO telescope. This implies that a balloon-born CHANT system would need a LIDAR with the laser pulse energy comparable to that of EUSO, to probe the atmosphere in the direction of each EAS candidate event. However, a space-borne CHANT system would observe EAS from the distance up to $2\times 10^3$~km, which is a factor 4 larger. This means that the required laser pulse energy should be an order-of-magnitude higher, compared to the EUSO or the balloon-borne CHANT. 

An alternative possibility for the space-borne CHANT system would be not to use the LIDAR for the atmospheric probe on event-by-event basis, but, instead, to  build a 3-d map of the optical properties of the atmosphere below the telescope. A dedicated LIDAR system able to scan the atmosphere over a 2000 km wide strip below the spacecraft, operating at a low Earth orbit, would provide a global 3d view of the cloud and aerosol coverage on a several orbit time scale (one orbit is approximately $T_{\rm orb}\simeq 90$~min) by exploiting the Earth rotation. The strip monitored at each subsequent orbit is shifted by $2\pi r_\oplus \left(T_{\rm orb}/24\mbox{ hr}\right)\simeq 2500$~km so that the strips monitored at subsequent orbits overlap. Scanning of a 2000 km wide strip  could be done using a LIDAR equipped with a laser pointing system, and with a wide field-of-view infrared camera, as it is done in JEM-EUSO telescope \cite{euso_ams}.

\subsection{Effective area}

The neutrino induced UAS are observable from an altitude $H$ in the distance range (see Fig. \ref{fig:tilt_mode} for geometrical notations)
\begin{equation}
R_{\rm uas}<R<R_{\rm hor} \,,
\end{equation}
where
\begin{equation}
R_{\rm uas}\simeq \frac{1}{2}\left[ \sqrt{8Hr_\oplus +\lambda_\nu^2}-\lambda_{\nu}\right]
\end{equation}
is the distance to the UAS emerging at maximal elevation angles.

The Cherenkov flux from an UAS is beamed in the forward direction. In the absence of scattering by aerosols, the characteristic beaming angle is  $\alpha\sim \alpha_{\rm Ch}\simeq 1.4^\circ$, about the Cherenkov angle in the atmosphere (Troposphere). An UAS which barely triggers the telescope camera is detectable only if the telescope is situated within the ``Cherenkov pool'' of the shower, a circle of  area 
\begin{eqnarray}
A_{\rm geom} & = & \pi \alpha_{\rm Ch}^2R^2 \nonumber \\
& \simeq & 2\times 10^3\left[\frac{R}{10^3\mbox{ km}}\right]^2\left[\frac{\alpha_{\rm Ch}}{1.5^\circ}\right]^2\mbox{ km}^2 \, .
\end{eqnarray}
Showers with energies well above the energy threshold could produce a detectable signal even if the observation angle is $\alpha>\alpha_{\rm Ch}$, because the angular distribution of Cherenkov light has tails  interpolating between a power-law and an exponential function  outside the Cherenkov cone $\alpha\gtrsim \alpha_{\rm Ch}$~\cite{wide_field_telescope,cherenkov_lateral_profile}. 

The UAS start in the atmosphere anywhere within the decay length of the tau lepton $\lambda_\tau$. For strongly inclined UAS, with  elevation angle  $\theta_E\sim 5^\circ$, the starting point of the shower is always within $\lesssim 1$~km from the ground. This lowest layer of the atmosphere  is characterised  by the presence of aerosols. Typical vertical depth of the night time aerosols is $\tau_{\rm aerosol}(\theta_E=\pi/2)\simeq 0.1$ over the ocean and twice as much over land~\cite{aerosol}.  The scale height of the aerosol layer is about \mbox{$h_{\rm aerosol}\sim 1$~km~\citep{aerosol}.}  Strongly inclined showers suffer from significant scattering on aerosols, with the optical depth reaching $\tau_{\rm aerosol}/\sin\theta_E\sim 1$.  The scattering phase function of aerosols peaks in the forward direction, with a typical opening angle $\alpha_{\rm aerosol}\sim 10^\circ$ \citep{aerosol_phase_function}. Scattering by aerosols redistributes a fraction $\{1-\exp[-\tau_{\rm aerosol}(\theta_E)]\}$ of the UAS Cherenkov light homogeneously within a cone with opening angle $\alpha\sim \alpha_{\rm aerosol}$. The combined effect of the broadening of the viewing angle due to the aerosol scattering and lateral profile of Cherenkov light could be approximated  by an ``effective'' UAS viewing angle, which scales with energy as
\begin{equation}
\alpha_{\rm eff}\sim \alpha_{\rm Ch}\left(\frac{E}{E_{\rm thr}}\right)^{\kappa} \,,
\end{equation} 
where $\kappa$ is a scaling power that could be derived from a detailed numerical calculation. 
For a detector which is fully efficient at $E_{\rm thr}$ (i.e., UAS at the distances up to $R_{\rm hor}$ trigger the telescope), the geometric area grows with energy as
\begin{eqnarray}
A_{\rm geom} & = & \pi\alpha_{\rm eff}^2R^2 \nonumber \\
& \simeq  & 1.5\times 10^3\left[\frac{H}{300\mbox{ km}}\right]\left[\frac{E}{E_{\rm thr}}\right]^{2\kappa}\mbox{ km}^2 \, . 
\end{eqnarray}
\begin{figure}
\includegraphics[width=\linewidth]{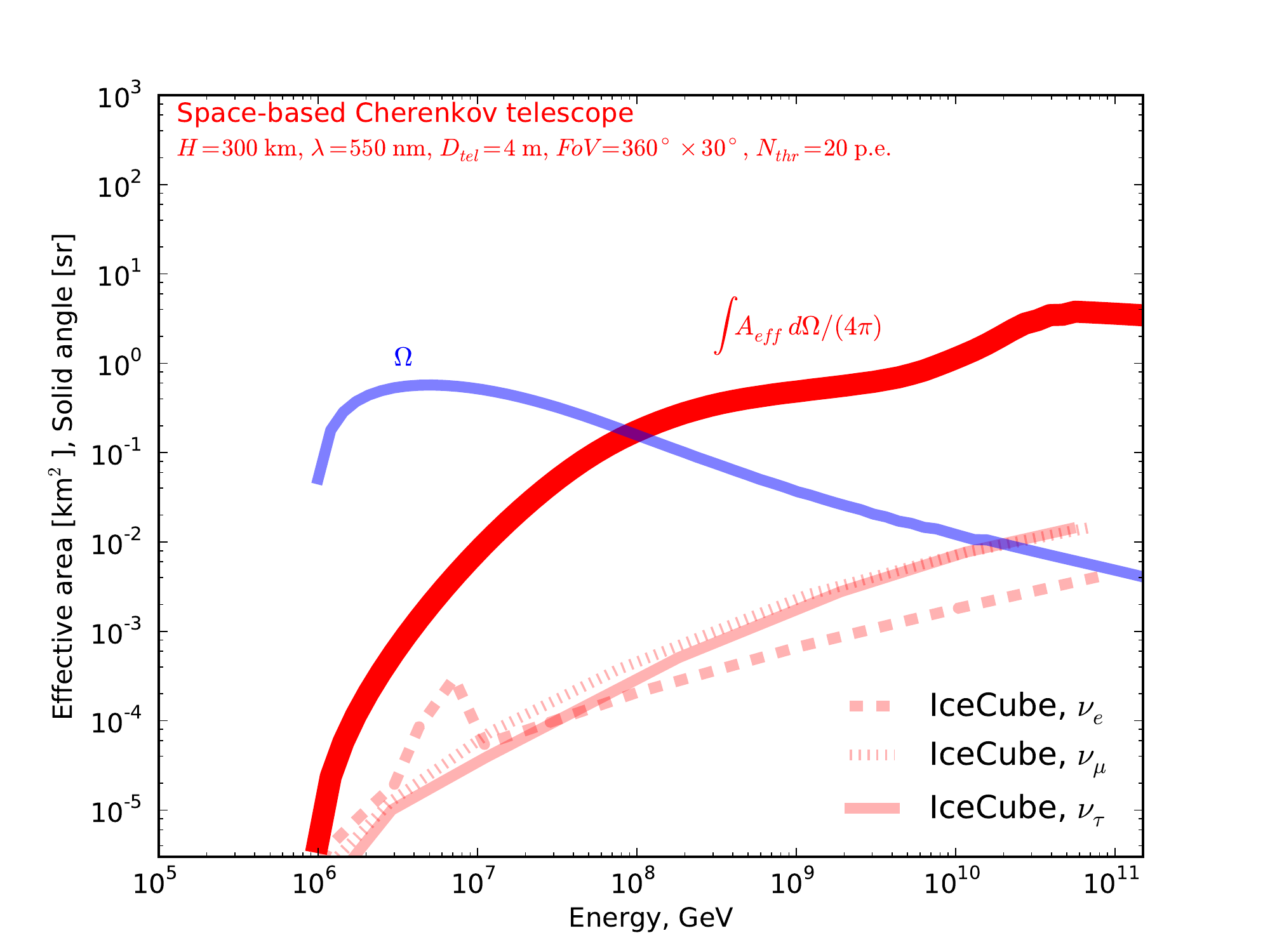}
\caption{Comparison of direction-averaged effective area of a space-based CHANT in the $\nu_\tau$ detection mode with the effective area of IceCube telescope \cite{IceCube_ICRC}. The blue curve shows the available observation solid angle ($\Omega\le 4\pi$) as a function of energy.}
\label{fig:aeff}
\end{figure}

Not all the neutrinos passing through the geometric area $A_{\rm geom}$ would produce a detectable UAS. The efficiency of conversion of neutrino into an UAS is determined by the ratio of the maximal possible depth of neutrino interactions  to the neutrino mean free path
\begin{equation}
p_{\nu_\tau}=\frac{\min(l_{\tau},\lambda_\tau)}{\lambda_\nu} \, .
\end{equation}
Substituting $\lambda_\tau$ from (\ref{eq:lambda}) we find that in the energy range $E<10^{18}$~eV a numerical estimate for $p_{\nu_\tau}$ is 
\begin{equation}
p_{\nu_\tau}\simeq 
3\times 10^{-3}
\left[\frac{E}{10^{17}\mbox{ eV}}\right]^{1.3} \, .
\end{equation}
For neutrinos with energies well above $10^{18}$~eV, the probability of  interaction resulting in an upgoing EAS is 
\begin{equation}
p_{\nu_\tau}\simeq
2\times 10^{-3}\left[\frac{E}{10^{17}\mbox{ eV}}\right]^{0.3} \, .
\end{equation}

The product
\begin{equation}
A_{\rm eff}=p_{\nu}A_{\rm geom}
\end{equation}
determines the effective area for neutrino detection. The result of a numerical  calculation of the effective area averaged over the solid angle $\Omega$,
\begin{equation}
\left<A_{\rm eff}\right>=\frac{1}{4\pi}\int A_{\rm eff} \   d\Omega \,,
\end{equation}
 is shown in Fig.~\ref{fig:aeff}. We have considered a reference telescope system consisting of a set of six refractor telescopes with diameter $D_{\rm tel}=4$~m and optical efficiency $\chi=0.1$, placed at an altitude $H=300$~km and overlooking a $30^\circ$ wide strip below the Earth horizon. In this calculation we have assumed that an UAS triggers the telescope if it produces an image of the size $N_{\rm thr}=20$ photoelectrons. We assume that the tau lepton receives on average 80\% of neutrino energy and its energy decreases by a factor of 3 after  the energy loss length given by Eq. (\ref{eq:ltau}). We also take into account the fact that the electromagnetic part of the UAS contains only about 50\% of the tau lepton energy if the decay is through the hadronic channel. 

Figure~\ref{fig:aeff} also shows the solid angle available for neutrino observations, as a function of energy. 
The energy dependence of $\Omega$ could be readily understood. At low energies, the rapid decrease of $\Omega$ is a threshold effect: the UAS are not bright enough to be detected by the telescope even at the largest elevation angle observable from the telescope position.  On the other hand, $\Omega$ decreases at  high energy because of the obscuration by the Earth. The maximum  value of $\Omega$ occurs at the energy threshold.

\subsection{Energy threshold  and energy resolution}

The threshold of a downward looking IACT could be estimated from the known thresholds of the upward pointed  IACTs. Assuming similar trigger conditions for the upward and downward looking IACTs (tens of photoelectrons contributing to the EAS image) we find a limit
\begin{eqnarray}
E>E_{\rm thr} & \simeq & 10\left[\exp(\tau_{\rm Rayleigh}+\tau_{\rm aerosol})\right]\left[\frac{D_{\rm tel}}{4\mbox{ m}}\right]^{-2}\nonumber\\ & \times &\left[\frac{R}{1000\mbox{ km}}\right]^{-2}\mbox{ PeV} \, ,
\label{eq:thr}
\end{eqnarray}
where $D_{\rm tel}$ is the diameter of the telescope dish (for a reflector) or entrance window (for a refractor telescope). This estimate reproduces the threshold of, e.g., the HEGRA-like telescopes FACT, and the Small-Size Telescopes of CTA, which have the dishes with 4~m diameter and will observe showers developing at a distance $R\sim 10$~km, comparable to the scale height of the atmosphere. For such setups, the above equation gives a threshold $E_{\rm thr}\sim 1$~TeV.

The molecular scattering  optical depth of the atmosphere $\tau_{\rm Rayleigh}$ is wavelength dependent:  
\begin{equation}
\tau_{\rm Rayleigh}\simeq 1.6\left[\frac{\lambda}{500\mbox{ nm}}\right]^{-4}\left[\frac{R_{\rm atm}}{100\mbox{ km}}\right] \,,
\end{equation}
where $R_{\rm atm}$ is the part of the path Cherenkov photons spend in the atmosphere below  its exponential scale height of about 8~km. The exponential factor in Eq.~(\ref{eq:thr}) grows rapidly with the decrease of the elevation angle of  the shower. For the distance range of interest, $R_{\rm atm}\sim 10^2$~km (valid for elevation angles $\theta_E\sim 5^\circ$),  the atmosphere is not transparent for the UV and blue light. The photosensors of the IACT should, therefore, be sensitive in the visible wavelength range to allow detection of the signal. 

\begin{figure}
\includegraphics[width=\linewidth]{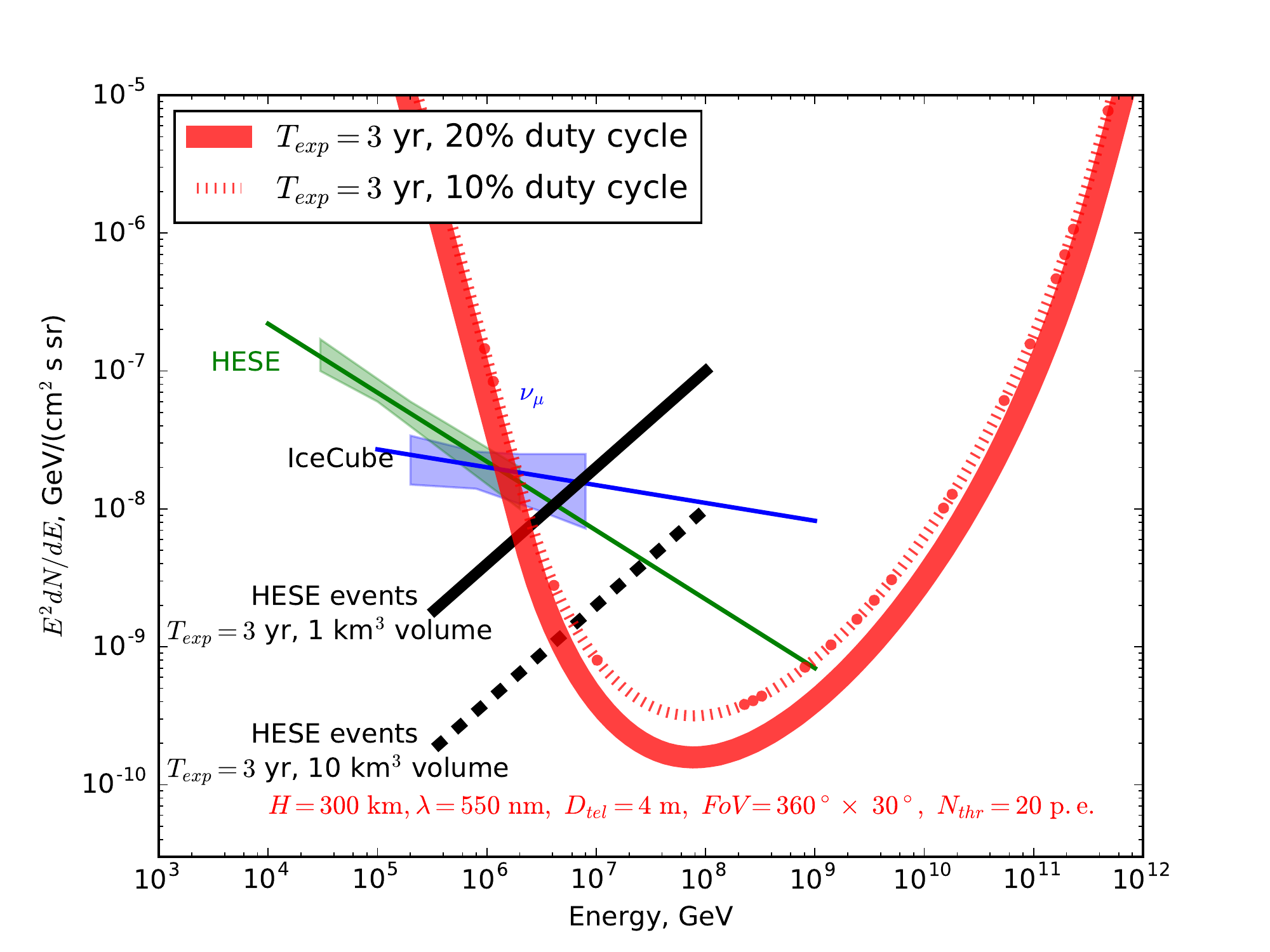}
\includegraphics[width=\linewidth]{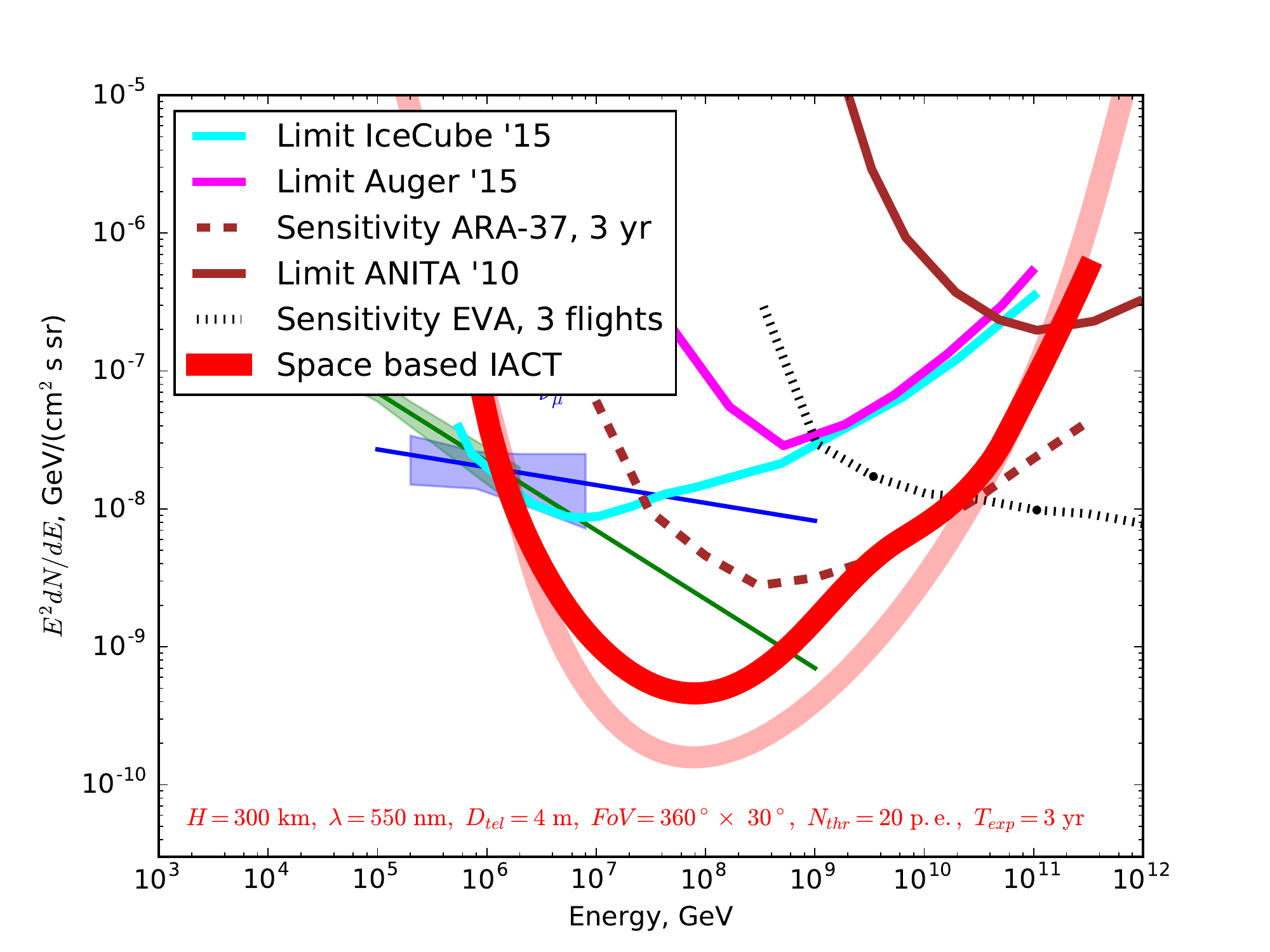}
\caption{IceCube measurement of the astrophysical neutrino flux in HESE (green range and line) and muon (blue range and line) neutrino channels \cite{,IceCube_combined_2015,icecube_6yr_muon} compared to sensitivity of three-year exposure of the space-based CHANT.  The top panel shows the sensitivity curve (red) defined as an envelope of the detection limits of power-law spectra with different slopes (grey lines) \cite{aguilar13}.  The bottom panel shows the quasi-differential sensitivity, defined by the requirement of detection of one event per decade of energy \cite{IceCube_ICRC}. The cyan and magenta curves show the upper limits from IceCube \cite{IceCube_ICRC} and the Pierre Auger Observatory \cite{auger15}. The black solid and dashed lines show  analytical sensitivity estimates for HESE events in $1$~km$^3$ and 10~km$^3$ ice- or water-Cherenkov detectors.  }
\label{fig:sensitivity}
\end{figure}

The rate of emission of Cherenkov photons by electrons in the Troposphere is estimated from the Tamm-Frank formula
\begin{eqnarray}
\frac{dN}{dl} & = & 2\pi\alpha(n^2-1)\frac{\Delta\lambda}{\lambda^2} \nonumber \\
& \simeq & 10\left[\frac{\lambda}{550\mbox{ nm}}\right]^{-2}\left[\frac{\Delta\lambda}{100\mbox{ nm}}\right]\frac{1}{\mbox{m}}
\end{eqnarray} 
where $n$ is the refraction index of the air and $\alpha$ is the fine structure constant. 
The number of Cherenkov radiation producing electrons in an EAS initiated by an interaction of a tau neutrino of the energy $E$ producing a $\tau$ lepton of comparable energy, which subsequently decays in the atmosphere before suffering significant energy loss is estimated as $N_e\sim 10^8\left[E/10^{17}\mbox{ eV}\right]$. GeV electrons lose energy on the Bremsstrahlung / ionisation loss distance scale $l_{e}\lesssim 10^3$~m so that every electron produces $\sim 10^4$ Cherenkov photons and the whole EAS produces
\begin{equation}
N_{\rm Ch}\sim 10^{12}\left[\frac{E}{10^{17}\mbox{ eV}}\right]\left[\frac{\lambda}{550\mbox{ nm}}\right]^{-2}\left[\frac{\Delta\lambda}{100\mbox{ nm}}\right]
\end{equation}
Cherenkov photons. Observation from a distance $R\sim 10^3\mbox{ km}$  with a telescope of diameter $D_{\rm tel}$ with the optical efficiency $\epsilon$ provides the photon statistics of the EAS image
\begin{eqnarray}
N & \sim & \epsilon N_{\rm Ch}e^{-\tau_{\rm Rayleigh}}\frac{(D_{tel}/R)^2}{4\alpha_{\rm Ch}^2} \nonumber \\ & \simeq & 6\times 10^2 e^{-\tau_{\rm Rayleigh}} \left[\frac{\epsilon}{0.1}\right] \left[\frac{D_{\rm tel}}{4\mbox{ m}}\right]^2\left[\frac{R}{10^3\mbox{ km}}\right]^{-2}\left[\frac{E}{10^{17}\mbox{ eV}}\right]\nonumber\\
&\times &\left[\frac{\lambda}{550\mbox{ nm}}\right]^{-2}\left[\frac{\Delta\lambda}{100\mbox{ nm}}\right] \,,
\end{eqnarray}
where $\alpha_{\rm Ch}\simeq 1.5^\circ$ is the Cherenkov angle in the Troposphere.

Apart from the Rayleigh scattering, the UAS also suffers from scattering on aerosols. As it is discussed above, the aerosols could have both harming and beneficial influence on the UAS signal. The harming effect is the reduction of direct Cherenkov light coming from the UAS  to the telescope. The beneficial effect is the spread of the Cherenkov photon flux into a wider cone with opening angle of about $10^\circ$. This improves the visibility of the UAS at larger off-axis angles and  leads to the increase of the effective collection area of the telescope system at  energies well  above the threshold. These two effects are taken into account in the numerical calculation used to derive the effective area and sensitivities in this paper. The ``bump'' in effective area in the $10^{19}$~eV energy range (Fig. \ref{fig:aeff}) is caused by the ``beneficial'' side of the aerosol influence. Contrary to the Rayleigh scattering, the wavelength dependence of the scattering on aerosol particles is weaker,  typically $\lambda^{-p_{\rm aerosol}}$, with the value of exponent $p_{\rm aerosol}$ determined by the distribution of sizes of the aerosol particles. For our calculations, we adopt the scaling $p_{\rm aerosol}=1$ so that  
\begin{equation}
\tau_{\rm aerosol}\simeq 0.5\left[\frac{\lambda}{500\mbox{ nm}}\right]^{-1}\left[\frac{R_{\rm aerosol}}{10\mbox{ km}}\right] \,,
\end{equation}
where $R_{\rm aerosol}$ is the path length of the UAS through the aerosol layer, which is calculated assuming the scale height of the aerosol layer $h_{\rm aerosol}\simeq 1$~km \cite{aerosol}.

The numerical code used for Fig. \ref{fig:aeff} and for the subsequent figures in the text takes into account both the molecular and aerosol scattering. The effective area shown in Fig. \ref{fig:aeff} is calculated for the reference wavelength $\lambda=550$~nm.

The energy resolution of the CHANT system is different in the energy ranges below and above $10^{17}$~eV, i.e., the energy at which $\lambda_\tau\simeq l_\tau$; see Eqs.~(\ref{eq:lambda}) and  (\ref{eq:ltau}). If the $\tau$ lepton decay distance $\lambda_\tau$ is shorter than the energy loss distance $l_\tau$, then the $\tau$ lepton which decays in the atmosphere retains most ($\sim 80\%$) of the energy of the primary $\nu_\tau$. The ``size'' of the upgoing EAS (the number of Cherenkov photons $N$) is proportional to the energy of the $\tau$ lepton, and so provides a measurement of the energy of the neutrino. The energy resolution reachable with the Cherenkov telescope techniques is down to $10-20\%$ \cite{hillas}.  It depends on the accuracy of reconstruction of the event geometry, on the statistics of the Cherenkov signal,  with a statistical error $\Delta E/E\simeq N^{-1/2}$, and on the statistical fluctuations of the EAS development.  

The energy resolution degrades in the energy range $E\gg 10^{17}$~eV in which the $\tau$ lepton suffers from the energy loss before decaying in the atmosphere. This poor energy resolution is a generic problem for all ultra-high-energy tau neutrino detection techniques. Measurement of the energy of the $\tau$ lepton based on the size of the EAS event provides only a lower bound on the energy of the primary $\nu_\tau$.

\subsection{Minimal detectable neutrino flux}

The minimal  detectable flux in an exposure $T_{\rm exp}$ is estimated as $E^2 dN_\nu/dE\sim 3E/(A_{\rm eff}T_{\rm exp}\Omega)$, where the factor of 3 accounts for the fact that only tau neutrinos are contributing to the detectable signal. These neutrinos constitute $1/3$ of the total neutrino flux under the standard  assumption of neutrino   production   in pion decays and maximal mixing between the muon and tau neutrino flavours.  Substituting the numerical estimates for $A_{\rm eff}$ and $\Omega$ computed in the previous section, one finds an estimate for the minimal detectable flux 
\begin{eqnarray}
E^2\frac{dN_\nu}{dE} & \sim & \frac{3E}{A_{\rm eff}T_{\rm exp}\Omega}\simeq 10^{-9}\left[\frac{E}{10^{16}\mbox{ eV}}\right]^{0.3-2\kappa}\nonumber\\ &\times &\left[\frac{H}{300\mbox{ km}}\right]^{-0.5}\left[\frac{T_{\rm exp}}{3\mbox{ yr}}\right]^{-1}\frac{\mbox{ GeV}}{\mbox{cm}^2\mbox{ s sr}} \, .
\end{eqnarray}

A more detailed numerical calculation of the minimal detectable flux, which  takes into account the change of the regime of propagation of the $\tau$ lepton above $10^{17}$~eV, is shown in Fig.~\ref{fig:sensitivity}. The two panels of the figure show two different definitions of the sensitivity limit.

The astrophysical neutrino fluxes in different models (Galactic, extragalactic diffuse background generated by AGN, etc.) predict broad energy distributions of neutrinos, well approximated by power-laws over many decades of energy. A useful definition of the sensitivity limit is the minimal detectable flux calculated under the assumption of a power-law neutrino spectrum. The reference slope of the power-law adopted in a large number of publications is $dN_\nu/dE\propto E^{-\Gamma}$ with $\Gamma=2$. However, models of the PeV-EeV astrophysical neutrino flux also predict $\Gamma$ different from $2$. The top panel of Fig.~\ref{fig:sensitivity} shows the sensitivity limit calculated for power-law spectra with arbitrary slopes. The red curve in the figure is an envelope curve for different minimal detectable power-law fluxes, as defined in Ref.  \cite{aguilar13}. For any $\Gamma$, the minimal detectable power-law spectrum is the straight line (examples are grey lines in the figure) tangent to the curve.

For each value of $\Gamma$, the minimal normalisation of the power-law flux is calculated from the requirement that the flux produces one detectable neutrino event in three years of operation, assuming a $20\%$ duty cycle. The duty cycle estimate includes the account of the astronomical night time at moderate moonlight conditions \cite{euso_astropart}. Presence of optically thick clouds which prevent observations of the UAS from above further reduces the efficiency of detection of the UAS and reduces the observation duty cycle by  a factor $\simeq 1...2$, depending on the efficiency of control of the cloud parameters with a dedicated atmospheric monitoring system~\cite{jemeuso_cloudy}. The convenience of this way of presentation of the sensitivity limit is  evident from a comparison of the power-law extrapolation of the measured astrophysical neutrino flux, assuming the slopes $\Gamma=2.5$ and $\Gamma=2.2$ (HESE and muon neutrino channel measurements by IceCube),  with the sensitivity limits for these slopes. This shows that  the extrapolated astrophysical fluxes are a factor of 10-100 higher than the sensitivity limits. This means that such flux levels are expected to give 10-100~events within a three year exposure, with the reference CHANT system.    

The bottom panel of Fig.~\ref{fig:sensitivity} shows another presentation of the sensitivity limit. In this representation, the neutrino flux at an energy $E$ is assumed to be a power-law with the slope $E^{-2}$ extending over an energy interval of fixed logarithmic width (one decade in energy) around  $E$. The minimal detectable flux is required to produce one neutrino event within a given exposure time. This definition of sensitivity is adopted in Refs.  \cite{IceCube_ICRC,auger15} which report bounds on the PeV-EeV band neutrino fluxes. 

Comparing this ``differential'' sensitivity with the ``integral'' sensitivity for the power-law type spectra one  can see that the integral sensitivity is better by a factor of 2 to 3  than the differential sensitivity in the energy range around $10^{17}-10^{18}$~eV. The sensitivity of the space-based CHANT is up to two  orders of magnitude better than the existing limits on neutrino flux in the energy range above 10~PeV \cite{IceCube_ICRC,auger15}, and is better than the sensitivity of ARA-37 {and ARIANNA} radio detection arrays \cite{ARA,ARIANNA}.

The sensitivity depends on a number of parameters of the CHANT system, such as the shower detection threshold, wavelength range, altitude, size of the FoV. The dependence of the sensitivity on these parameters is discussed in Appendix~\ref{appA}.

\subsection{Background }

There are three types of events which could occasionally be misinterpreted as neutrino induced UAS. First, random coincidences of excesses of the night photon background counts might occasionally exceed the detection threshold of the telescope camera. Another source of background is the signal from cosmic ray induced air showers. Finally, shower-like signals could be produced by the direct interactions of cosmic rays with the telescope and the focal plane detector.

\subsubsection{Night Earth Background }

A way to suppress the night Earth background  (NEB) is to take only events with sufficiently high number of photoelectrons simultaneously triggering many adjacent pixels. This background suppression method is  used in the astronomical observations of \gr\ sources with IACTs.  Pixels of the telescope perceive the flux from the night Earth  atmosphere with the counting rate {\cite{airglow,TUS,airglow_ref}}
\begin{eqnarray}
{\cal R} & \sim  & 10^7\left[\frac{\epsilon}{0.1}\right]\left[\frac{D_{\rm tel}}{4\mbox{ m}}\right]^2\left[\frac{\Omega_{\rm uas}}{0.02~{\rm deg}^2}\right]\left[\frac{\Delta\lambda}{100\mbox{ nm}}\right] \nonumber \\
& \times & \left[\frac{\lambda}{400\mbox{ nm}}\right]\frac{1}{\mbox{ s}} \,,
\label{eq:back}
\end{eqnarray}
where $\epsilon\simeq 0.1$ is the product of the optics throughput and photon detection efficiency of the photosensor, $D_{\rm tel}$ is the diameter of the telescope, and $\Omega_{\rm uas}$ is the solid angle spanned by the UAS image. The NEB is a combination of the atmospheric airglow and scattered emission from the starlight. The airglow rate scales roughly proportionally to $\lambda$ in the wavelength range of interest \cite{airglow, airglow_ref}.   

{The brightness of the airglow component of the NEB depends on the position of the telescope pixel in the FoV. Pixels close to the Earth horizon direction perceive higher airglow flux due to the limb-brightening effect. The airglow layer is situated in the Thermosphere in the altitude range above $\sim 100$~km. This makes the limb-brighening effect rather moderate for the directions below the Earth horizon. It is taken into account in the numerical calculation of the sensitivity curves.}

We assume that part of the FoV affected by city lights is excluded from the exposure area as it is done for the space-based UHECR detector JEM-EUSO \cite{jemeuso}. Pixels affected by strong city lights will be excluded based on the high background rate above a certain threshold, measured in the ``slow mode'' of operation of the telescopes (static photos taken at a fixed rate). Pixels affected by a moderate level human-induced light pollution below the threshold will still be used, with the EAS detection threshold adjusted according to the background rate level.  {The level of the background created by the city lights could be estimated based on the V band measurements reported in the Ref. \citep{citylights}. The vertical emissivity of artificial lights varies from below $10^{7}$~ph/cm$^2$s~sr at the outskirts of the metropolitan areas to more than $10^{9}$~ph/cm$^2$s~sr at locations of big cities. Observed from the distance $\sim R_{\rm uas}$ the flux perceived by a group of pixels spanning the solid angle $\Omega_{\rm uas}$ and overlooking an area occupied by a city of the size $l=D_{\rm uas}\sqrt{\Omega_{\rm uas}}\sim 1 \dots 10$~km occasionally found in the FoV is   
\begin{eqnarray}
{\cal R}_{\rm V,city} & \sim  & 10^6 \dots 10^{8}\left[\frac{\epsilon}{0.1}\right]\left[\frac{D_{\rm tel}}{4\mbox{ m}}\right]^2\left[\frac{\Omega_{\rm uas}}{0.02~{\rm deg}^2}\right]
\label{eq:city}
\end{eqnarray}
in the V band at $\lambda\simeq 550$~nm, $\Delta\lambda\simeq 90$~nm (and assuming that the city light emission is isotropic). The artificial background could exceed the airglow-related background by more than order of magnitude in relatively large city areas. This could produce chains of false triggers, if no special arrangement to increase the detection threshold in the pixels exposed to city lights is foreseen. Taking this into account, }
the level of the natural and human-made background light has to be continuously monitored on different time scales, because of the transient nature of some of the human-produced backgrounds. This would require a dedicated ``slow-mode'' of CHANT operation (similar to that of the EUSO telescope \cite{euso_slow_mode}) in which photographs of the monitored region are taken at a rate higher than the typical time scale of variations of the human made backgrounds.

The NEB rate has also a contribution from the scattered moonlight. Strong moonlight limits the observation duty cycle. The experience of operation of ground-based Cherenkov telescopes shows that observations at different moonlight conditions are possible, especially with the Silicon photomultipliers which do not experience degradation at high illumination rates \cite{fact1}. Only a small several day period around the full Moon {and the direction of observations around the Earth surface projection of the line connecting the Moon to the telescope}  have to be avoided. We adopt an estimate of the 20\% duty cycle which includes the scattered moonlight effect based on Ref. \cite{euso_astropart}.  

{Still another type of the signal from the night Earth could be produced by the lightning and the associated phenomena in thunderstorm regions (elves, jets etc). These phenomena occur on longer, millisecond, time scale. A dedicated trigger has to be set up for those phenomena, to avoid damaging of photo-detectors and to exclude the time intervals and pixels affected by the thunderstorms from the neutrino exposure areas. The strategy  could be identical to that implemented for JEM-EUSO \cite{jets}.} 

{Complexity of the multi-component NEB described above indicates that a detailed observational study of the NEB should be a subject of a dedicated R\&D project, e.g. a dedicated balloon flight over representative areas such as, sea, scaresly populated ground, densely populated areas etc., before the full-scale CHANT mission.}  

The NEB induced triggers could be efficiently suppressed by suitable adjustment of the trigger threshold. In the simplest version, the threshold could be imposed on the minimal number of photoelectrons in the shower image, $N_{\rm thr}$.

The probability to find a fluctuation of the count rate of the night photon background comparable to the signal of an UAS, e.g., $N_{\rm thr}=20$ photoelectrons within the time interval  $\delta t\sim 50$~ns in {spread between several adjacent pixels spanning the solid angle  $\Omega_{\rm uas}$}, is $p\sim 10^{-26}$ for the average count rate given in Eq.~(\ref{eq:back}).  Overall, the focal plane of the telescope assembly with the FoV $360^\circ\times 30^\circ$ contains about $N\sim 3\times 10^5$ different two-three pixel spots. The probability to find a fluctuation of the NEB above $N_{\rm thr}$ anywhere in the FoV is $Np\sim 3\times 10^{-21}$ in each  time interval $\delta t$. Observations on the time span of $T_{\rm exp}\sim 10^3$~days, with a duty cycle $\simeq 0.2$, provide about $T\sim 2\times 10^7$~seconds of observation time, or about $T/\delta t\sim 10^{15}$ possible time slots. The probability for a fluctuation above $N_{\rm thr}$ to occur is therefore $NpT/\delta t\sim 10^{-5}$. Thus, choosing the detection threshold at about $\sim 20$~photoelectrons in several adjacent pixels  should be sufficient for the suppression of  the random coincidence background for a particular choice of telescope parameters considered in the example. Note that the images of distant UAS detectable by space-based CHANT are typically very compact, spanning just one-to-several pixels in the focal plane (assuming the pixel size is about $0.1^\circ$). This is different from the ground-based IACTs, where the shower images are much larger. Compactness of the image reduces the NEB for the image detection (compared to the night sky background in the case of ground-based IACT). 

\subsubsection{UHECR event background}

Another type of background for neutrino detection is generated by UHECR events. UHECR induced EAS produce Cherenkov and fluorescence light which might be mis-interpreted as the UAS Cherenkov light. 

The fluorescence and scattered Cherenkov emission from the UHECR EAS tracks could be readily distinguished from the UAS Cherenkov signal based on different timing properties of the two types of signal. The Cherenkov emission from the UAS is beamed in the direction of the telescope, so that the signal is contained within a short time interval, typically within $\Delta t\lesssim 10-50$~ns. To the contrary, typical time scale of the UHECR signal is determined by the time of transit of the EAS through the atmospheric column of about 10~km. This takes $\gtrsim 30\ \mu$s, a time scale which is longer by a factor of 100. 

A faster signal from UHECR induced UAS is coming from the  ``Cherenkov mark'' of the EAS on the ground. This mark is produced by the Cherenkov light scattered at the footprint of the shower on the ground (or on a cloud). The footprint spans several hundred meters on the ground and the time scale of the Cherenkov mark is at least a factor of 10 shorter than the time scale of the fluorescence signal from the EAS track in the atmosphere. The Cherenkov mark time scale could occasionally be as short as the UAS signal time scale. 

Contrary to the direct Cherenkov light from an UAS, the emission from the Cherenkov mark of an UHECR shower is isotropized in the $4\pi$ solid angle. This means that the intensity of the scattered light from the footprint is typically $\pi \alpha_{\rm Ch}^2/(4\pi)\sim 10^{-4}$ times weaker than that of the direct Cherenkov light. Cosmic rays which could trigger the downward looking IACTs should have energies at least $10^4$ times larger than the energies of the tau leptons initiating the detectable UAS. Taking into account that the energy threshold for the UAS detection is in the 10~PeV range, the showers with detectable Cherenkov marks should originate from UHECR with energies about $10^{20}$~eV. 

The UHECR shower Cherenkov mark background for the UAS detection could be suppressed via identification of the fluorescence track component of the same UHECR showers. Indeed, a shower which produces a strong Cherenkov mark containing more than $N_{\rm thr}$ photoelectrons would typically have much more photoelectrons constituting an image of the EAS track \cite{jemeuso_da}. This is true also for the showers with particularly strong Cherenkov marks on optically thick clouds \cite{jemeuso_cloudy}. As it is mentioned above the fluorescence track image is formed on much longer time scale of $\Delta T\gtrsim 30\ \mu$s. The NEB accumulated on this time scale has statistics ${\cal R}\Delta T$ by a factor of $100$ higher, compared to the background within a $\delta t$ window of the UAS Cherenkov signal detection. The fluorescence track is detectable if the statistics of photoelectrons in the track exceeds the background fluctuations, $\sqrt{{\cal R}\Delta T}$. The threshold for the UAS detection could be adjusted in such a way that  the UHECR Cherenkov mark events could be efficiently rejected based on the detection of the fluorescence track features. For the particular choice of parameters of the reference CHANT system, this requirement gives an adjusted threshold value close to that needed to suppress the NEB. 

Still another tool for discrimination of the UAS and UHECR Cherenkov mark events is available in the case of the UAS in which the tau lepton decays via hadronic channel (approximately 60\% of all the events). These UAS have a significant muon component. Low energy (1-10 GeV) muons in the UAS lose energy via ionisation, $\sim 2$~MeV/(g/cm$^2)$, on the distance range $\sim 100$~km. Longer tracks of muons (compared to the electromagnetic shower component)  produce UAS tracks resolvable with a telescope of $0.1^\circ$ resolution, even from distances $\sim 10^3$~km \cite{muon_showers}. The time scale on which the muon trace of the UAS could be seen is much shorter than that of the fluorescence signal from the UHECR shower tracks,  because of the strong Doppler beaming of the signal. This time scale difference provides a discrimination between the fluorescence tracks of UHECR EAS and the muon Cherenkov tracks of the UAS events.  

Events produced by neutrinos with energies in the UHECR range occur in a very thin layer close to the Earth limb. In this region a higher cosmic ray event background is produced by the horizontal EAS initiated by cosmic rays with nearly horizontal incidence angles. This background could be rejected in a straightforward way, by retaining only the events which develop below the Earth horizon. However, limited angular resolution of the telescope would not allow rejection of the cosmic ray events in a narrow strip of the width about the telescope point spread function, $\theta_{PSF}\simeq 0.1^\circ$. We take this into account by imposing a veto on events occurring in this strip. This reduces CHANT sensitivity in the energy range above $10^{19}$~eV.   

\subsubsection{Background from cosmic rays hitting the detector}

Still another type of events which could mimic the UAS signals could be produced by interactions of cosmic rays with the detector. A cosmic ray passing through the photosensor would induce a particle cascade which might produce a signal above the detection threshold $N_{\rm thr}$. Such a signal might even be spread over several pixels if the cosmic ray energy is high enough and the incidence angle is large enough. This type of events could be rejected based on the morphology of the image. The event induced by a cosmic ray hitting the detector has a ``single pixel'' or a ``track'' morphology, depending on the parameters of the cosmic ray. Events produced by cosmic rays hitting other parts of the telescope (lens and baffle) would produce events with morphology of diffuse wide ``spots.'' To the contrary, an event produced by an UAS has imaging characteristics determined by the shape of the telescope point spread function. For example, the EUSO refractive optics produces a PSF with a characteristic vignetting pattern and this is typical for any wide field-of-view optics \cite{euso_optics}.    Detailed analysis of the background induced by cosmic rays hitting the telescope would require  simulations which take into account the detailed geometry of the system, choice of materials etc. It is outside the scope of this paper. 

\section{Discussion}

We have demonstrated that observations of UAS with the space-based CHANT can detect astrophysical neutrinos with energies above 10 PeV. Development of the new technique of observations of UAS with IACT systems can complement existing neutrino detection techniques which run out of signal statistics above several PeV energy. To see the complementarity explicitly,  it is useful to directly compare the sensitivities of the two techniques directly.

\subsection{Comparison with the sensitivity of water and ice Cherenkov detectors to the neutrino flux}

The established technique of detection of the atmospheric and astrophysical neutrinos is  via sampling of Cherenkov light from the  tracks or showers of charged particles produced via charged current interactions of neutrinos.  The IceCube telescope detects the Cherenkov radiation using a km$^3$ scale network of photomultipliers buried in Antarctic ice. 

A first determination of the sensitivity of IceCube like detectors  could be found from an estimate of its effective area, which is a product of the geometric area of the detector $A_{\rm geom}\simeq 1$~km$^2$ times the efficiency of detection of neutrinos passing through the detector volume.  The neutrino mean free path in ice or water is \mbox{$\lambda_{\nu}\simeq  4.3\times 10^{8}
\left[E/10^{17}\mbox{ eV}\right]^{-0.3}\mbox{ cm}
$.} Comparing this mean free path to the thickness of the detector $\sim 1$~km, one could find the efficiency of neutrino detection as a function of energy, $p_\nu\sim 2\times 10^{-4} \left[E/10^{17}\mbox{ eV}\right]^{0.3}$. The IceCube effective area for the HESE type signal is then
\begin{equation}
A_{\rm eff}=p_\nu \, A_{\rm geom}\simeq 2\times 10^{-4}\left[\frac{E}{10^{17}\mbox{ eV}}\right]^{0.3}\mbox{ km}^2 \, .
\end{equation}
The observation solid angle of the HESE events is macroscopically large at all energies, extending to nearly the whole sky below  PeV energy range and shrinking to less than $2\pi$ at higher energies, that is in the energy range in which the Earth is not transparent to neutrinos.  Figure \ref{fig:sensitivity} shows an estimate of the IceCube sensitivity for HESE events assuming $2\pi$ angular acceptance. {This estimate does not take into account the loss of efficiency of the detector for events with energies close to the detection threshold. It also does not take into account the deviation of the energy dependence of the neutrino interaction cross-seciton from the powerlaw. The esitimate is valid only for the HESE events. The through-going muon event detection channel is characterised by larger effective area at high energies, but it is also characterised by the smaller angular acceptance, due to the opacity of the Earth. Account of the thoughoing muon detection channel results in higher sensitivity at the highest energies, as shown in Fig. \ref{fig:sensitivity_cosmogenic}.} 

A ten-fold increase of the detector volume planned for the IceCube Generation 2 facility will result in an order-of-magnitude improvement of $A_{\rm eff}$ and, as a result, of sensitivity for the HESE events. However, one could deduce from Fig. \ref{fig:sensitivity} that even in this case the sensitivity of the IACT array is better than the ice-Cherenkov detector in the energy range above $10^{16}$~eV. 

The superior sensitivity of the space-based IACT is crucial for the study of the spectrum of the astrophysical signal and, in particular, for the detection of a possible high-energy cut-off of astrophysical neutrino spectrum. Such a cut-off is expected for the Galactic component of the neutrino flux, because our Galaxy could hardly host sources capable of producing cosmic rays with energies up to $10^{18}$~eV. If this is so, cosmic ray interactions in the sources and in the interstellar medium could not produce neutrinos with energies higher than $\sim 10^{16}$~eV. 

One more important advantage of CHANT is the precise angular resolution. Muon track features detectable in 60\% of the UAS events provide a possibility for high-precision angular reconstruction of the UAS arrival direction, using the standard technique of the IACT systems.   Typical precision of the reconstruction of directions of the EAS detected with IACTs is one-to-two orders of magnitude better than the angular resolution of the HESE signal in  IceCube. Precise localization of the arrival direction of individual neutrino events on the sky is not crucially important if the neutrino signal is diffuse in nature. However, if the signal has a contribution from isolated individual sources, precise localization of the source on the sky is important for the source identification with the methods of multi-wavelength astronomy.  In addition to multi-wavelength astronomy, localization of neutrino clusters in the sky is important to search for these extremely energetic sources through, e.g.,  correlation with bursting sources such as \gr\ bursts, AGN flares, compact object (black hole and neutron star) merger LIGO events, etc. CHANT will be able to provide a followup observations of the transient events on different time scales, depending on the configuration. A space-based telescope at low Earth orbit would scan the sky on the orbital time scale, $T_{\rm orb}\simeq 90$~min. The balloon and ground-based telescopes are able to do followup observations on $\simeq 1$~d time scale. {For comparison, a ground based detector IceCube detector, being situated as the South Pole, has a possibility to instantly follow up the transient events in the Southern sky in the HESE detection mode and of events occurring in a narrow strip around zero declination in the through-going muon event detection mode in the 10-100 PeV energy range.}

\subsection{Detection of the cosmogenic neutrino flux}

\begin{figure}
\includegraphics[width=\linewidth]{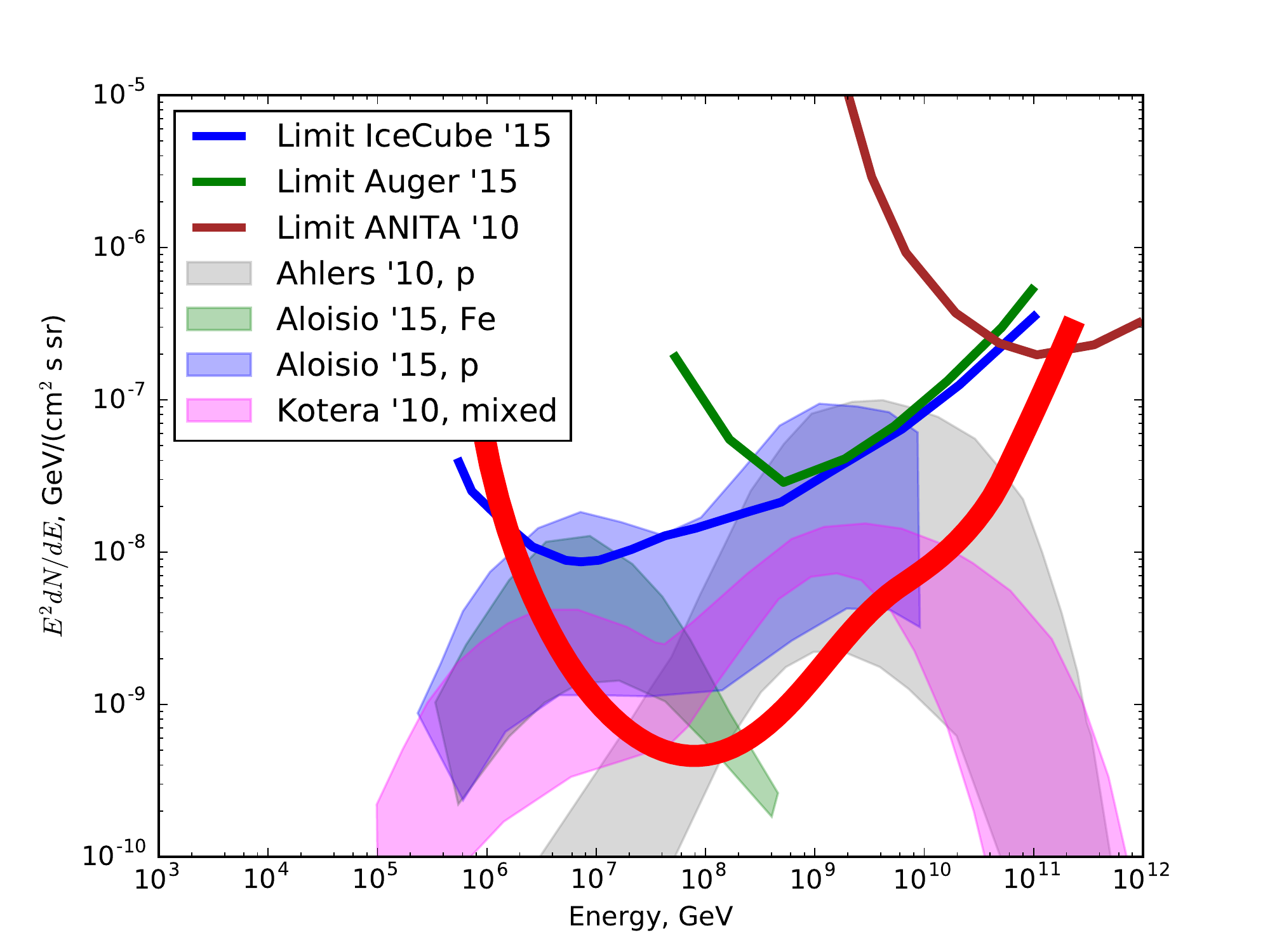}
\caption{Comparison of the sensitivity of space borne IACT system to the expected levels of cosmogenic neutrino flux, modelled under different assumptions about UHECR flux composition and radiation backgrounds in the intergalactic medium   \cite{ahlers10,kotera10,aloisio15}. Limits from Icecube \cite{IceCube_ICRC}, Auger \cite{auger15} and ANITA \cite{ANITA} are shown for comparison.}
\label{fig:sensitivity_cosmogenic}
\end{figure}

Another type of neutrino signal expected in the energy range above 10~PeV is the neutrinos produced in interactions of UHECR with the CMB and EBL. This signal has not yet been detected, but it is generically expected to exist. The strength of the signal depends on a number of factors. 

First of all, it is strongly dependent on the nature of the highest energy cosmic rays. If they are protons, a large cosmogenic neutrino signal is expected to appear in the EeV energy range due to the interactions of the UHECR protons with the CMB. Another component of the signal at lower energies is expected from interactions of UHECR protons with the photons of EBL, see Fig. \ref{fig:sensitivity_cosmogenic}.

The strength of the signal depends on the law of evolution of the UHECR source population. Model calculations shown in Fig. \ref{fig:sensitivity_cosmogenic} span a range of spectral shapes and normalisations determined by different assumptions about cosmological evolution of the UHECR source populations (evolution which follows the star formation rate,  or the AGN population, etc.). 

If most of the UHECR are heavy nuclei, the energy per nucleon is below the pion production threshold for interactions with the CMB. Only the UHECR-EBL interaction component of neutrino flux is present, as shown in Fig. \ref{fig:sensitivity_cosmogenic}. Most of the cosmogenic neutrino flux is in the 10~PeV energy band in this case. The strength of the neutrino signal also depends on the law of evolution of the UHECR source population. 

The sensitivity level achievable with the UAS observations by space-based CHANT is in principle sufficient for the exploration of the full parameter space of the UHECR induced neutrino signal. Models that minimise photopion production at the source, while reproducing the spectrum and mass composition of UHECRs yield a neutrino flux considerably smaller than those shown in Fig. 4 (see e.g., Fig. 12 in Ref.~\cite{Unger:2015laa}). Even these extremely low neutrino fluxes will be within the CHANT discovery reach. In particular, the non-detection of the higher-energy "bump" of the cosmogenic neutrino spectrum (Fig. \ref{fig:sensitivity_cosmogenic}) would firmly rule out the possibility that UHECR flux has a sizeable proton component.  
 
\subsection{Space vs. balloon vs. ground-based IACT systems}

The sensitivity of  CHANT to UAS depends on the altitude of the telescope. The maximal observable area scales proportionally to the square of the distance to the horizon and proportionally to the altitude $H$, see  Eq. (\ref{eq:r_hor}). Thus, lifting the telescope to space provides up to two orders of magnitude gain in the effective collection area and, as a result, two orders of magnitude gain in sensitivity. 

However, this gain is achievable only in the ``full efficiency'' regime in which the telescope triggers on UAS events occurring at the distances about $R_{\rm hor}$. Such UAS observed from higher and higher altitude  become more and more distant. This leads to the shift of the minimal energy of the full efficiency regime of IACT systems at higher altitudes toward higher energies. The resulting sensitivity gain of the higher altitude IACT system is then less than two orders of magnitude. 

\begin{figure}
\includegraphics[width=\linewidth]{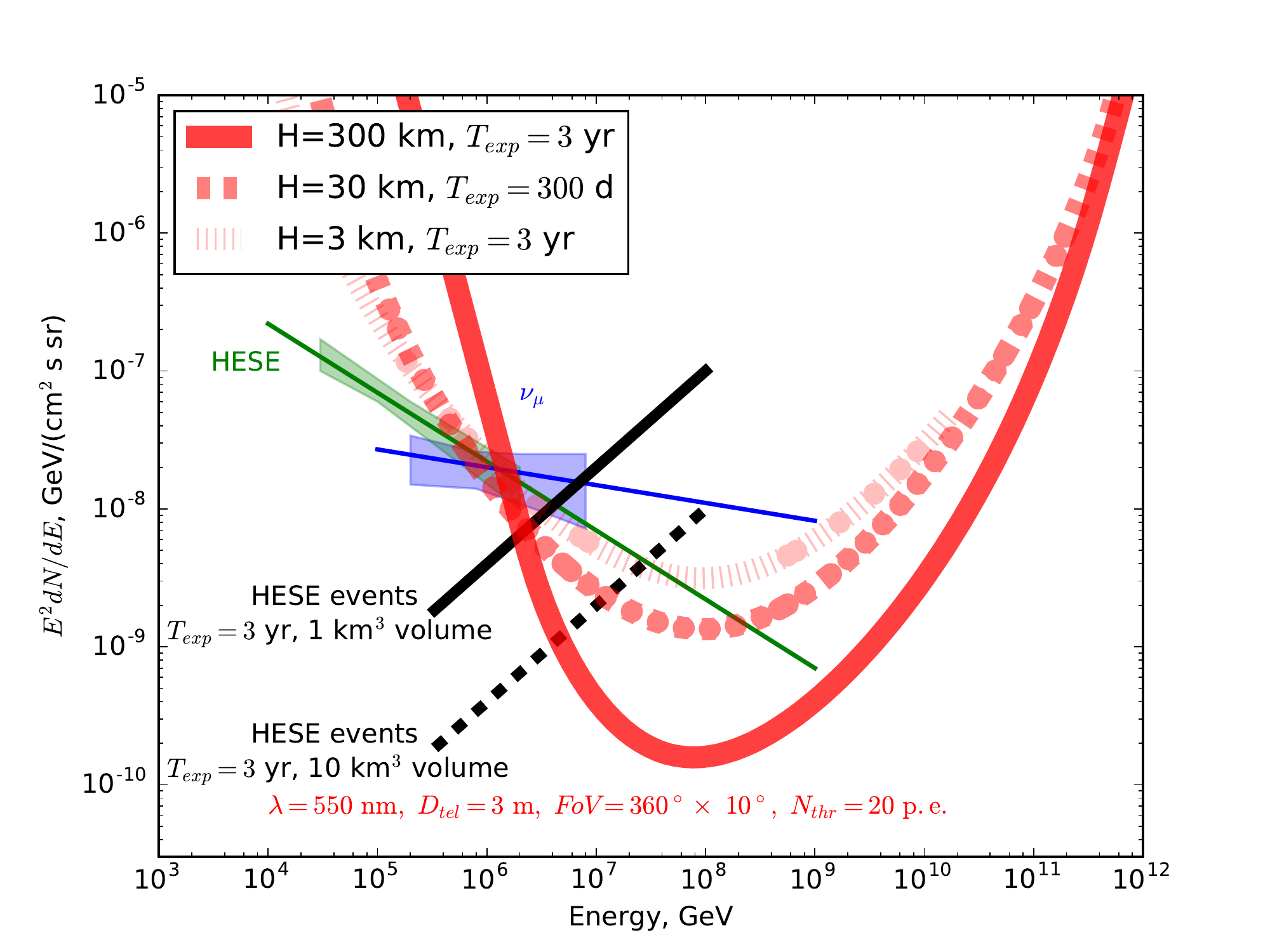}
\caption{Comparison of the integral sensitivity of the space borne IACT system to the balloon borne and ground-based systems with identical parameters. For the balloon, the 300~days exposure corresponds to three flight campaigns of long-duration SPB balloon flight.}
\label{fig:sensitivity_balloon}
\end{figure}

This is illustrated in Fig. \ref{fig:sensitivity_balloon} which compares the integral sensitivities of ground-based, balloon borne and space-borne  CHANT systems with identical characteristics. One could see that lifting the telescope to space provides more than an order-of-magnitude gain in sensitivity for an $E^{-2}$ type spectrum, compared to the ground-based system. 

A balloon borne CHANT system can be an order of magnitude more 
sensitive than a ground system. However, ground-systems can operate for longer periods.  The newly started Super Pressure Balloons (SPB) campaigns by NASA \citep{spb} has vastly expanded the available observation time for balloon payloads.   SPB technology enables flights for months. A sequence of  flight campaigns can provide 300~days exposure over the ocean from 40 km altitude with significant gains as compared to  a year long exposure of a ground base system. A CHANT-SPB  balloon payload should be able to prove the space-based technology and discover the cosmogenic neutrinos.

\acknowledgments We thank Michael Unger for valuable comments. LAA  is supported by U.S. National Science Foundation (NSF) CAREER Award PHY1053663 and by the National Aeronautics and Space Administration (NASA) grant  NNX13AH52G; he thanks the Center for Cosmology and Particle Physics at New York University for its hospitality. JHA is supportted by NASA grant NNX13AH53G. AVO acknowledges support from the NSF grant PHY-1412261 and the NASA grant NNX13AH54G at the University of Chicago, and the NSF grant PHY-1125897 at the Kavli Institute for Cosmological Physics.

\begin{figure}
\includegraphics[width=\linewidth]{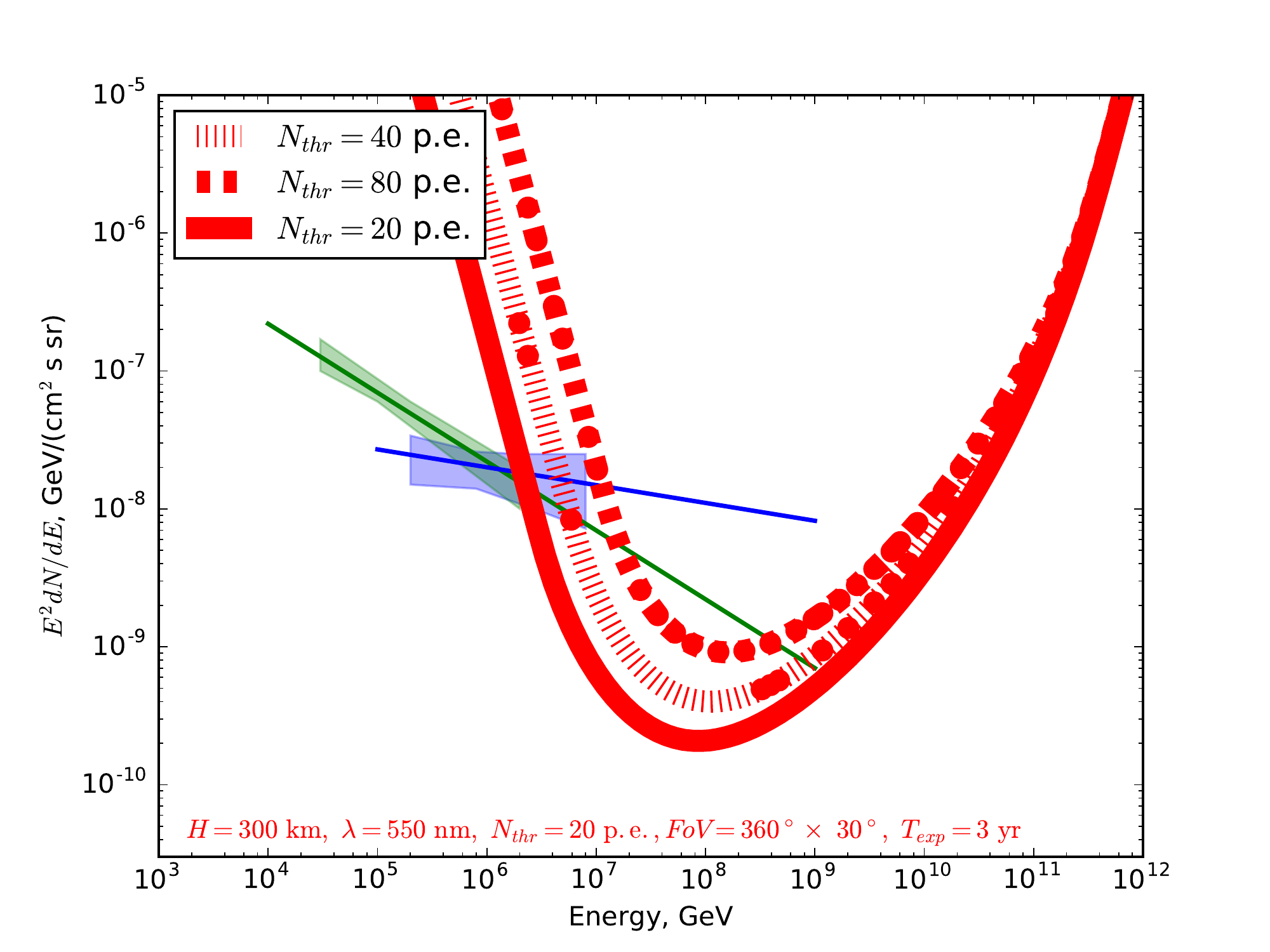}
\includegraphics[width=\linewidth]{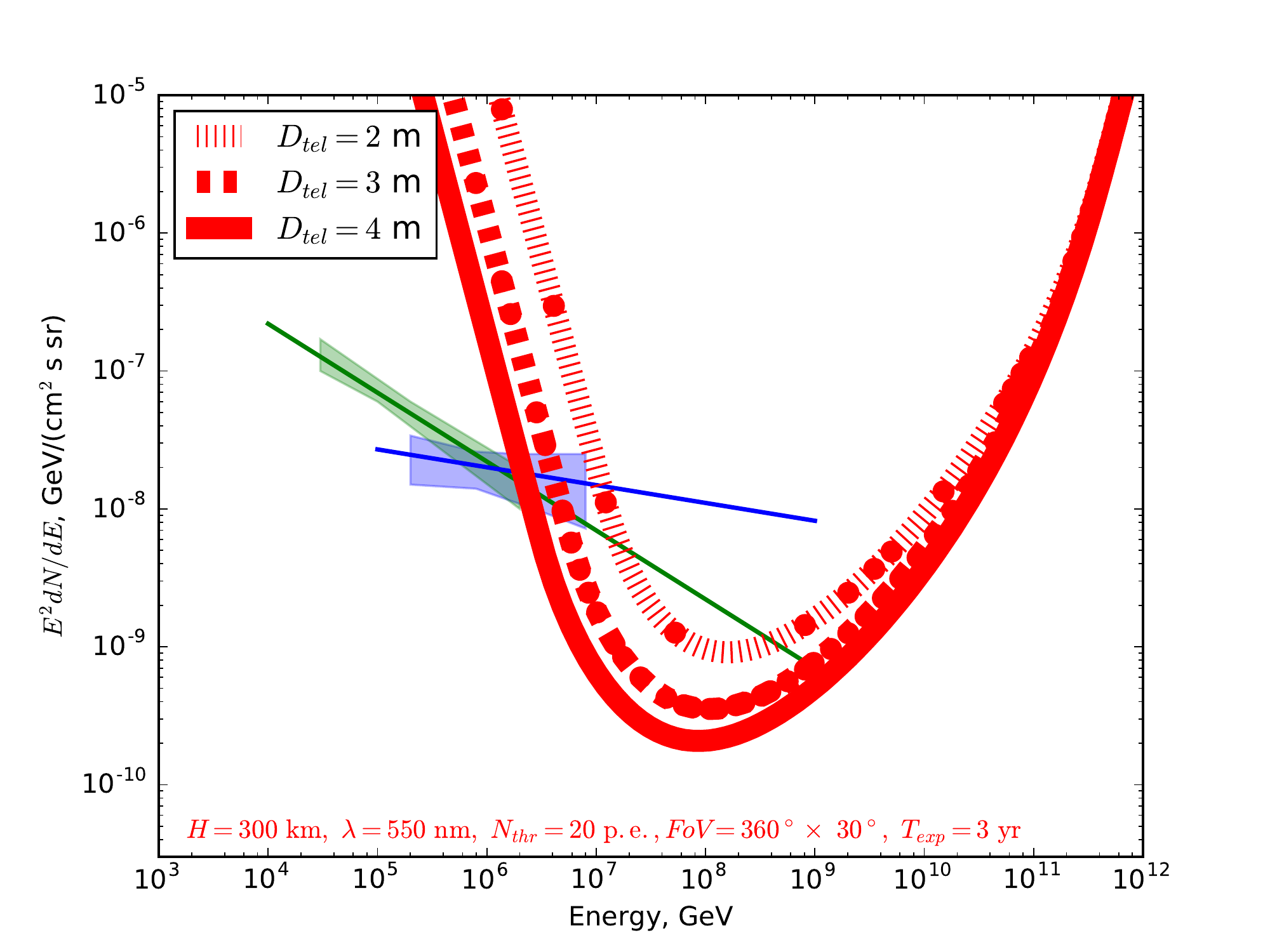}
\caption{Same as in Fig. \ref{fig:sensitivity}, but for different detection thresholds (top) and telescope sizes (bottom). }
\label{fig:sensitivity_dtel}
\end{figure}
\begin{figure}
\includegraphics[width=\linewidth]{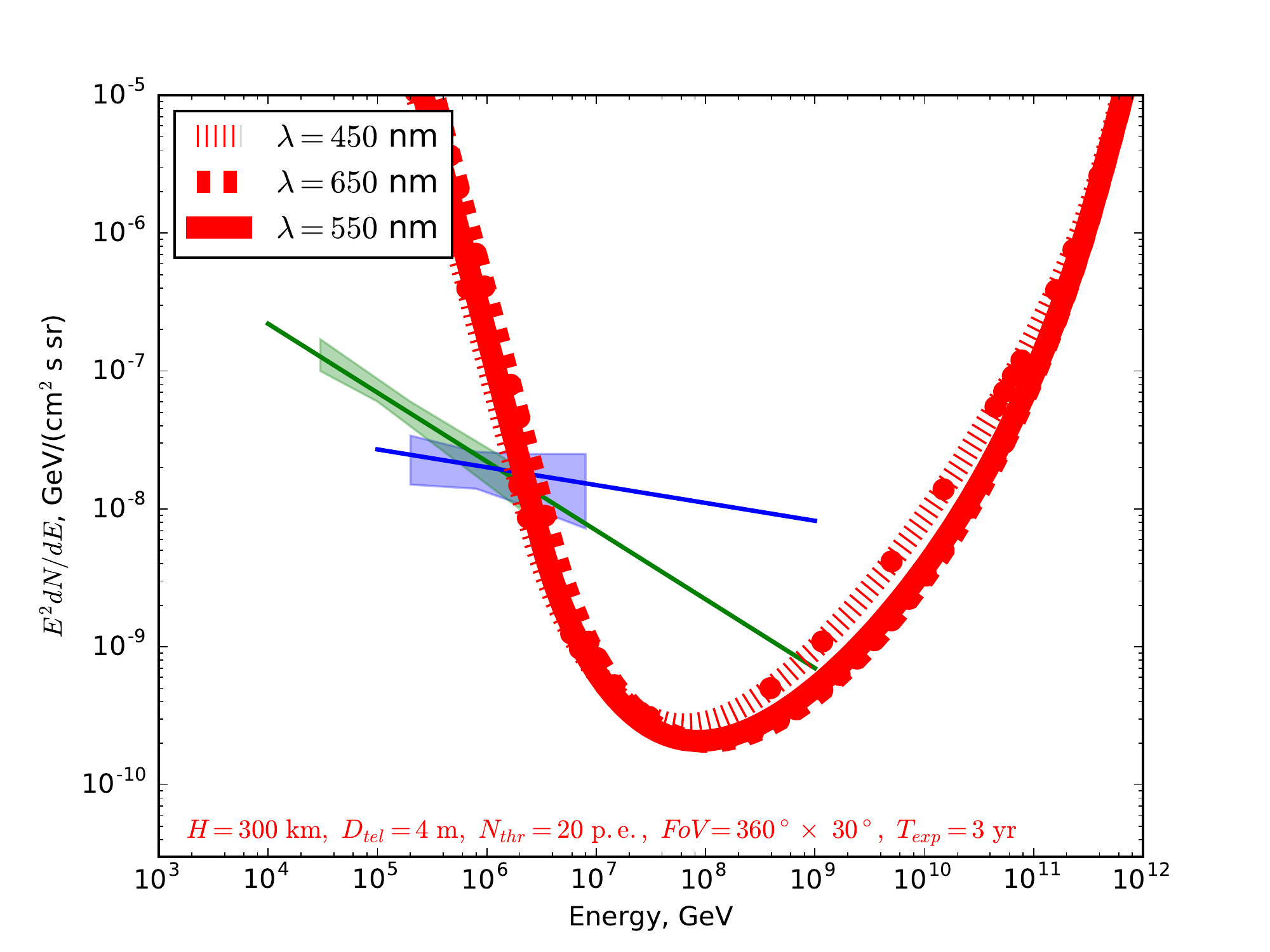}
\includegraphics[width=\linewidth]{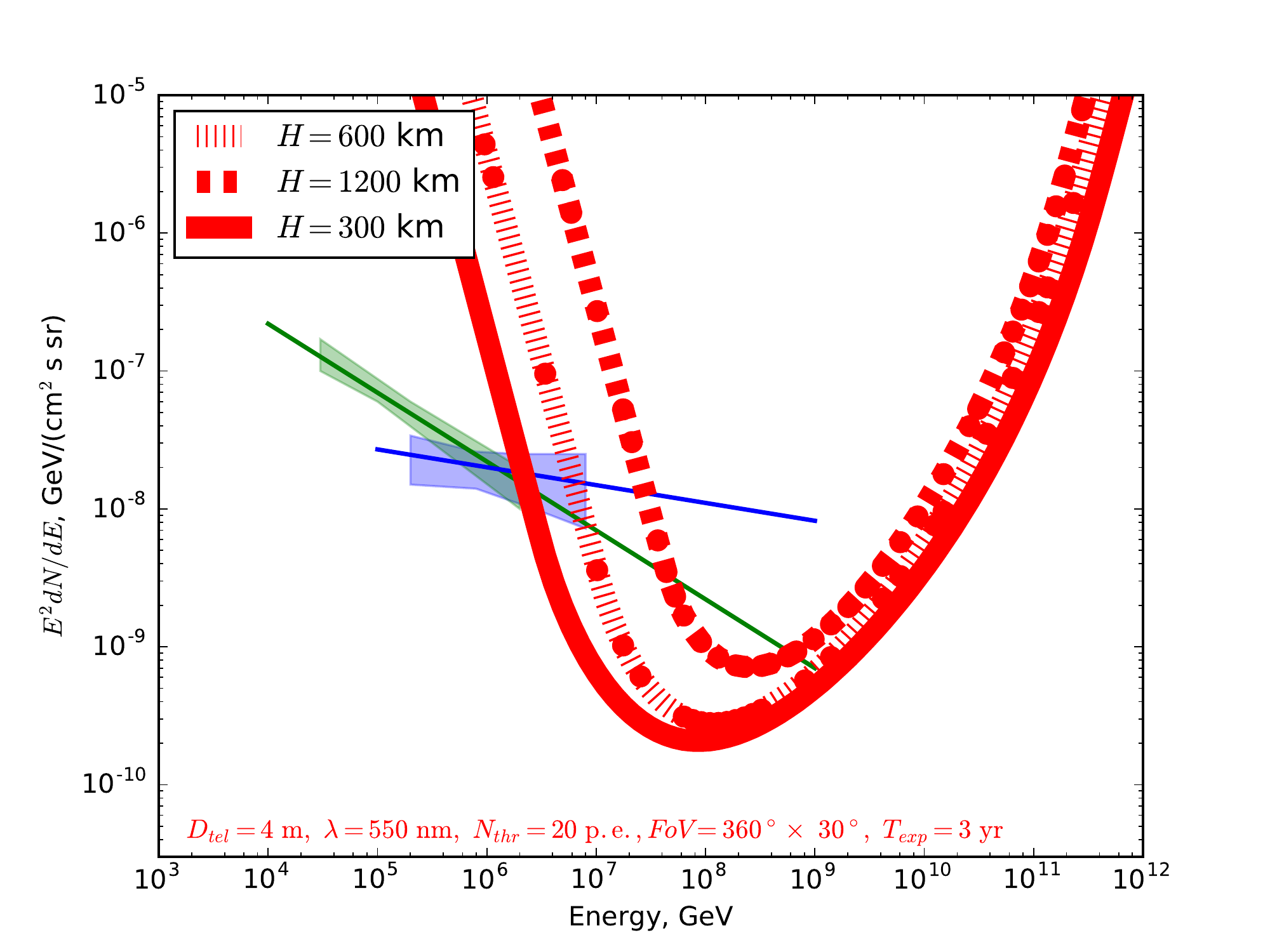}
\caption{Same as in Fig. \ref{fig:sensitivity}, but for different central wavelength of photo-sensors (top) and different flight altitudes (bottom).  }
\label{fig:sensitivity_lambda}
\end{figure}
\begin{figure}
\includegraphics[width=\linewidth]{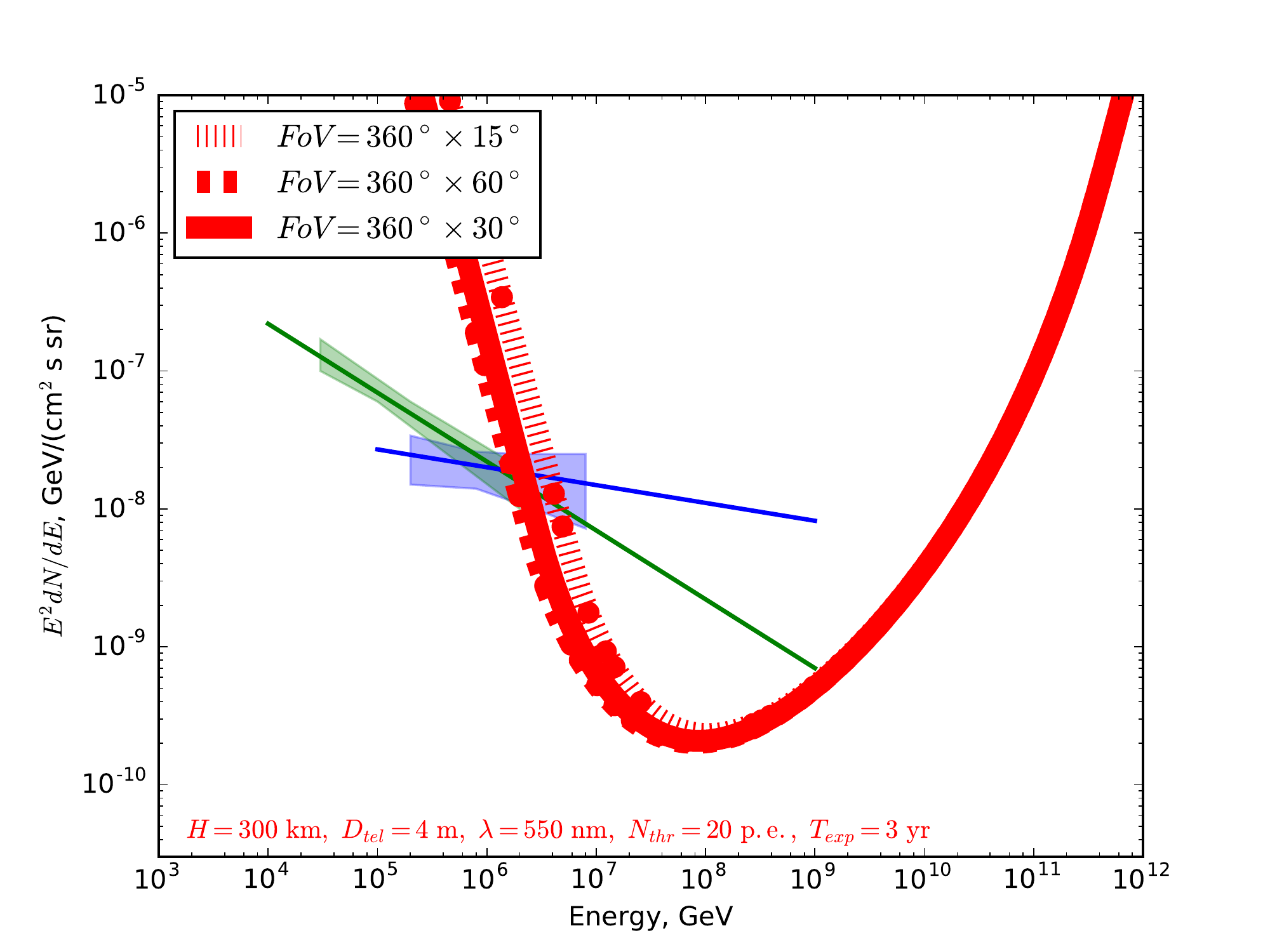}
\caption{Same as in Fig. \ref{fig:sensitivity}, but for different telescope FoVs.}
\label{fig:sensitivity_fov}
\end{figure}

\appendix
\section{Sensitivity dependence of space-based IACT system on telescope parameters}
\label{appA}

In this Appendix we study the dependence of the sensitivity on different parameters characterizing the triggering and identification of  neutrino events.  As a first exercise we vary the detector threshold and size. The results are encapsulated in Fig.~\ref{fig:sensitivity_dtel}. We can see that increasing the number of photoelectrons required for event triggering by factors of 2 and 3, roughly reduces the experimental sensitivity (at and below the peak) by the same factors. Increasing the diameter of the telescope by a factor a 2 leads to a sensitivity about a factor of 4 higher. In Fig.~\ref{fig:sensitivity_lambda}  we study the  dependence of the trigger with the  altitude of the orbit and with our fiducial wavelength: $\lambda = 550~{\rm nm}$. As can be seen by inspection, for energies below the sensitivity peak the detector performance is invariant under changes  of $\lambda$ in the range $450 \leq \lambda/{\rm nm} \leq 650$. For  $\lambda \alt 550~{\rm nm}$, there is a mild dependence that diminishes the sensitivity with rising energy.  Finally, in Fig.~\ref{fig:sensitivity_fov} we show that there is almost no variation with the FoV.

\bibliography{EUSO_neutrinos}

\begin{thebibliography}{88}
\expandafter\ifx\csname natexlab\endcsname\relax\def\natexlab#1{#1}\fi
\expandafter\ifx\csname bibnamefont\endcsname\relax
  \def\bibnamefont#1{#1}\fi
\expandafter\ifx\csname bibfnamefont\endcsname\relax
  \def\bibfnamefont#1{#1}\fi
\expandafter\ifx\csname citenamefont\endcsname\relax
  \def\citenamefont#1{#1}\fi
\expandafter\ifx\csname url\endcsname\relax
  \def\url#1{\texttt{#1}}\fi
\expandafter\ifx\csname urlprefix\endcsname\relax\def\urlprefix{URL }\fi
\providecommand{\bibinfo}[2]{#2}
\providecommand{\eprint}[2][]{\url{#2}}

\bibitem[{\citenamefont{{Aartsen}
  et~al.}(2015{\natexlab{a}})}]{IceCube_combined_2015}
\bibinfo{author}{\bibfnamefont{M.~G.} \bibnamefont{{Aartsen}}}
  \bibnamefont{et~al.}, \bibinfo{journal}{\apj} \textbf{\bibinfo{volume}{809}},
  \bibinfo{eid}{98} (\bibinfo{year}{2015}{\natexlab{a}}), \eprint{1507.03991}.

\bibitem[{\citenamefont{{Aartsen}
  et~al.}(2015{\natexlab{b}})}]{IceCube_muonnu_2015}
\bibinfo{author}{\bibfnamefont{M.~G.} \bibnamefont{{Aartsen}}}
  \bibnamefont{et~al.}, \bibinfo{journal}{Physical Review Letters}
  \textbf{\bibinfo{volume}{115}}, \bibinfo{eid}{081102}
  (\bibinfo{year}{2015}{\natexlab{b}}), \eprint{1507.04005}.

\bibitem[{\citenamefont{{The IceCube Collaboration}
  et~al.}(2015)\citenamefont{{The IceCube Collaboration}, {Aartsen}, {Abraham},
  {Ackermann}, {Adams}, {Aguilar}, {Ahlers}, {Ahrens}, {Altmann}, {Anderson}
  et~al.}}]{IceCube_ICRC}
\bibinfo{author}{\bibnamefont{{The IceCube Collaboration}}},
  \bibinfo{author}{\bibfnamefont{M.~G.} \bibnamefont{{Aartsen}}},
  \bibinfo{author}{\bibfnamefont{K.}~\bibnamefont{{Abraham}}},
  \bibinfo{author}{\bibfnamefont{M.}~\bibnamefont{{Ackermann}}},
  \bibinfo{author}{\bibfnamefont{J.}~\bibnamefont{{Adams}}},
  \bibinfo{author}{\bibfnamefont{J.~A.} \bibnamefont{{Aguilar}}},
  \bibinfo{author}{\bibfnamefont{M.}~\bibnamefont{{Ahlers}}},
  \bibinfo{author}{\bibfnamefont{M.}~\bibnamefont{{Ahrens}}},
  \bibinfo{author}{\bibfnamefont{D.}~\bibnamefont{{Altmann}}},
  \bibinfo{author}{\bibfnamefont{T.}~\bibnamefont{{Anderson}}},
  \bibnamefont{et~al.}, \bibinfo{journal}{ArXiv e-prints}
  (\bibinfo{year}{2015}), \eprint{1510.05223}.

\bibitem[{\citenamefont{{IceCube Collaboration}
  et~al.}(2016)\citenamefont{{IceCube Collaboration}, {Aartsen}, {Abraham},
  {Ackermann}, {Adams}, {Aguilar}, {Ahlers}, {Ahrens}, {Altmann}, {Andeen}
  et~al.}}]{icecube_6yr_muon}
\bibinfo{author}{\bibnamefont{{IceCube Collaboration}}},
  \bibinfo{author}{\bibfnamefont{M.~G.} \bibnamefont{{Aartsen}}},
  \bibinfo{author}{\bibfnamefont{K.}~\bibnamefont{{Abraham}}},
  \bibinfo{author}{\bibfnamefont{M.}~\bibnamefont{{Ackermann}}},
  \bibinfo{author}{\bibfnamefont{J.}~\bibnamefont{{Adams}}},
  \bibinfo{author}{\bibfnamefont{J.~A.} \bibnamefont{{Aguilar}}},
  \bibinfo{author}{\bibfnamefont{M.}~\bibnamefont{{Ahlers}}},
  \bibinfo{author}{\bibfnamefont{M.}~\bibnamefont{{Ahrens}}},
  \bibinfo{author}{\bibfnamefont{D.}~\bibnamefont{{Altmann}}},
  \bibinfo{author}{\bibfnamefont{K.}~\bibnamefont{{Andeen}}},
  \bibnamefont{et~al.}, \bibinfo{journal}{ArXiv e-prints}
  (\bibinfo{year}{2016}), \eprint{1607.08006}.

\bibitem[{\citenamefont{Aartsen et~al.}(2016)}]{Aartsen:2016ngq}
\bibinfo{author}{\bibfnamefont{M.~G.} \bibnamefont{Aartsen}}
  \bibnamefont{et~al.} (\bibinfo{collaboration}{IceCube})
  (\bibinfo{year}{2016}), \eprint{1607.05886}.

\bibitem[{\citenamefont{{Kistler} and {Laha}}(2016)}]{multi-PeV1}
\bibinfo{author}{\bibfnamefont{M.~D.} \bibnamefont{{Kistler}}}
  \bibnamefont{and} \bibinfo{author}{\bibfnamefont{R.}~\bibnamefont{{Laha}}},
  \bibinfo{journal}{ArXiv e-prints}  (\bibinfo{year}{2016}),
  \eprint{1605.08781}.

\bibitem[{\citenamefont{{Neronov} and {Semikoz}}(2016{\natexlab{a}})}]{neronov}
\bibinfo{author}{\bibfnamefont{A.}~\bibnamefont{{Neronov}}} \bibnamefont{and}
  \bibinfo{author}{\bibfnamefont{D.}~\bibnamefont{{Semikoz}}},
  \bibinfo{journal}{Astroparticle Physics} \textbf{\bibinfo{volume}{75}},
  \bibinfo{pages}{60} (\bibinfo{year}{2016}{\natexlab{a}}),
  \eprint{1509.03522}.

\bibitem[{\citenamefont{{Neronov} and
  {Semikoz}}(2016{\natexlab{b}})}]{neronov1}
\bibinfo{author}{\bibfnamefont{A.}~\bibnamefont{{Neronov}}} \bibnamefont{and}
  \bibinfo{author}{\bibfnamefont{D.}~\bibnamefont{{Semikoz}}},
  \bibinfo{journal}{Astroparticle Physics} \textbf{\bibinfo{volume}{72}},
  \bibinfo{pages}{32} (\bibinfo{year}{2016}{\natexlab{b}}), \eprint{1412.1690}.

\bibitem[{\citenamefont{Razzaque}(2013)}]{Razzaque:2013uoa}
\bibinfo{author}{\bibfnamefont{S.}~\bibnamefont{Razzaque}},
  \bibinfo{journal}{Phys. Rev.} \textbf{\bibinfo{volume}{D88}},
  \bibinfo{pages}{081302} (\bibinfo{year}{2013}), \eprint{1309.2756}.

\bibitem[{\citenamefont{Bai et~al.}(2014)\citenamefont{Bai, Barger, Barger, Lu,
  Peterson, and Salvado}}]{Bai:2014kba}
\bibinfo{author}{\bibfnamefont{Y.}~\bibnamefont{Bai}},
  \bibinfo{author}{\bibfnamefont{A.~J.} \bibnamefont{Barger}},
  \bibinfo{author}{\bibfnamefont{V.}~\bibnamefont{Barger}},
  \bibinfo{author}{\bibfnamefont{R.}~\bibnamefont{Lu}},
  \bibinfo{author}{\bibfnamefont{A.~D.} \bibnamefont{Peterson}},
  \bibnamefont{and} \bibinfo{author}{\bibfnamefont{J.}~\bibnamefont{Salvado}},
  \bibinfo{journal}{Phys. Rev.} \textbf{\bibinfo{volume}{D90}},
  \bibinfo{pages}{063012} (\bibinfo{year}{2014}), \eprint{1407.2243}.

\bibitem[{\citenamefont{Anchordoqui}(2016)}]{Anchordoqui:2016dcp}
\bibinfo{author}{\bibfnamefont{L.~A.} \bibnamefont{Anchordoqui}}
  (\bibinfo{year}{2016}), \eprint{1606.01816}.

\bibitem[{\citenamefont{{Neronov} and
  {Semikoz}}(2016{\natexlab{c}})}]{neronov_muon}
\bibinfo{author}{\bibfnamefont{A.}~\bibnamefont{{Neronov}}} \bibnamefont{and}
  \bibinfo{author}{\bibfnamefont{D.~V.} \bibnamefont{{Semikoz}}},
  \bibinfo{journal}{ArXiv e-prints}  (\bibinfo{year}{2016}{\natexlab{c}}),
  \eprint{1603.06733}.

\bibitem[{\citenamefont{Anchordoqui
  et~al.}(2014{\natexlab{a}})\citenamefont{Anchordoqui, Goldberg, Lynch,
  Olinto, Paul, and Weiler}}]{Anchordoqui:2013qsi}
\bibinfo{author}{\bibfnamefont{L.~A.} \bibnamefont{Anchordoqui}},
  \bibinfo{author}{\bibfnamefont{H.}~\bibnamefont{Goldberg}},
  \bibinfo{author}{\bibfnamefont{M.~H.} \bibnamefont{Lynch}},
  \bibinfo{author}{\bibfnamefont{A.~V.} \bibnamefont{Olinto}},
  \bibinfo{author}{\bibfnamefont{T.~C.} \bibnamefont{Paul}}, \bibnamefont{and}
  \bibinfo{author}{\bibfnamefont{T.~J.} \bibnamefont{Weiler}},
  \bibinfo{journal}{Phys. Rev.} \textbf{\bibinfo{volume}{D89}},
  \bibinfo{pages}{083003} (\bibinfo{year}{2014}{\natexlab{a}}),
  \eprint{1306.5021}.

\bibitem[{\citenamefont{{Ahlers} and {Halzen}}(2015)}]{ahlers_review}
\bibinfo{author}{\bibfnamefont{M.}~\bibnamefont{{Ahlers}}} \bibnamefont{and}
  \bibinfo{author}{\bibfnamefont{F.}~\bibnamefont{{Halzen}}},
  \bibinfo{journal}{Reports on Progress in Physics}
  \textbf{\bibinfo{volume}{78}}, \bibinfo{eid}{126901} (\bibinfo{year}{2015}).

\bibitem[{\citenamefont{{Giacinti} et~al.}(2015)\citenamefont{{Giacinti},
  {Kachelriess}, {Kalashev}, {Semikoz}, and {Neronov}}}]{giacinti15}
\bibinfo{author}{\bibfnamefont{G.}~\bibnamefont{{Giacinti}}},
  \bibinfo{author}{\bibfnamefont{M.}~\bibnamefont{{Kachelriess}}},
  \bibinfo{author}{\bibfnamefont{O.}~\bibnamefont{{Kalashev}}},
  \bibinfo{author}{\bibfnamefont{D.~V.} \bibnamefont{{Semikoz}}},
  \bibnamefont{and} \bibinfo{author}{\bibfnamefont{A.~N.}
  \bibnamefont{{Neronov}}}, \bibinfo{journal}{ArXiv e-prints}
  (\bibinfo{year}{2015}), \eprint{1507.07534}.

\bibitem[{\citenamefont{{Stecker} et~al.}(1991)\citenamefont{{Stecker}, {Done},
  {Salamon}, and {Sommers}}}]{Stecker91}
\bibinfo{author}{\bibfnamefont{F.~W.} \bibnamefont{{Stecker}}},
  \bibinfo{author}{\bibfnamefont{C.}~\bibnamefont{{Done}}},
  \bibinfo{author}{\bibfnamefont{M.~H.} \bibnamefont{{Salamon}}},
  \bibnamefont{and}
  \bibinfo{author}{\bibfnamefont{P.}~\bibnamefont{{Sommers}}},
  \bibinfo{journal}{Physical Review Letters} \textbf{\bibinfo{volume}{66}},
  \bibinfo{pages}{2697} (\bibinfo{year}{1991}).

\bibitem[{\citenamefont{{Mannheim} and {Biermann}}(1992)}]{mannheim92}
\bibinfo{author}{\bibfnamefont{K.}~\bibnamefont{{Mannheim}}} \bibnamefont{and}
  \bibinfo{author}{\bibfnamefont{P.~L.} \bibnamefont{{Biermann}}},
  \bibinfo{journal}{\aap} \textbf{\bibinfo{volume}{253}}, \bibinfo{pages}{L21}
  (\bibinfo{year}{1992}).

\bibitem[{\citenamefont{{Neronov} and {Semikoz}}(2002)}]{neronov_semikoz02}
\bibinfo{author}{\bibfnamefont{A.~Y.} \bibnamefont{{Neronov}}}
  \bibnamefont{and} \bibinfo{author}{\bibfnamefont{D.~V.}
  \bibnamefont{{Semikoz}}}, \bibinfo{journal}{\prd}
  \textbf{\bibinfo{volume}{66}}, \bibinfo{eid}{123003} (\bibinfo{year}{2002}),
  \eprint{hep-ph/0208248}.

\bibitem[{\citenamefont{{Tchernin}
  et~al.}(2013{\natexlab{a}})\citenamefont{{Tchernin}, {Aguilar}, {Neronov},
  and {Montaruli}}}]{tchernin13}
\bibinfo{author}{\bibfnamefont{C.}~\bibnamefont{{Tchernin}}},
  \bibinfo{author}{\bibfnamefont{J.~A.} \bibnamefont{{Aguilar}}},
  \bibinfo{author}{\bibfnamefont{A.}~\bibnamefont{{Neronov}}},
  \bibnamefont{and}
  \bibinfo{author}{\bibfnamefont{T.}~\bibnamefont{{Montaruli}}},
  \bibinfo{journal}{\aap} \textbf{\bibinfo{volume}{555}}, \bibinfo{eid}{A70}
  (\bibinfo{year}{2013}{\natexlab{a}}), \eprint{1305.3524}.

\bibitem[{\citenamefont{{Loeb} and {Waxman}}(2006)}]{loeb06}
\bibinfo{author}{\bibfnamefont{A.}~\bibnamefont{{Loeb}}} \bibnamefont{and}
  \bibinfo{author}{\bibfnamefont{E.}~\bibnamefont{{Waxman}}},
  \bibinfo{journal}{\jcap} \textbf{\bibinfo{volume}{5}}, \bibinfo{eid}{003}
  (\bibinfo{year}{2006}), \eprint{astro-ph/0601695}.

\bibitem[{\citenamefont{{Murase} et~al.}(2013)\citenamefont{{Murase}, {Ahlers},
  and {Lacki}}}]{murase}
\bibinfo{author}{\bibfnamefont{K.}~\bibnamefont{{Murase}}},
  \bibinfo{author}{\bibfnamefont{M.}~\bibnamefont{{Ahlers}}}, \bibnamefont{and}
  \bibinfo{author}{\bibfnamefont{B.~C.} \bibnamefont{{Lacki}}},
  \bibinfo{journal}{\prd} \textbf{\bibinfo{volume}{88}}, \bibinfo{eid}{121301}
  (\bibinfo{year}{2013}), \eprint{1306.3417}.

\bibitem[{\citenamefont{{Tamborra} et~al.}(2014)\citenamefont{{Tamborra},
  {Ando}, and {Murase}}}]{tamborra}
\bibinfo{author}{\bibfnamefont{I.}~\bibnamefont{{Tamborra}}},
  \bibinfo{author}{\bibfnamefont{S.}~\bibnamefont{{Ando}}}, \bibnamefont{and}
  \bibinfo{author}{\bibfnamefont{K.}~\bibnamefont{{Murase}}},
  \bibinfo{journal}{\jcap} \textbf{\bibinfo{volume}{9}}, \bibinfo{eid}{043}
  (\bibinfo{year}{2014}), \eprint{1404.1189}.

\bibitem[{\citenamefont{{Bechtol} et~al.}(2015)\citenamefont{{Bechtol},
  {Ahlers}, {Di Mauro}, {Ajello}, and {Vandenbroucke}}}]{no_starforming}
\bibinfo{author}{\bibfnamefont{K.}~\bibnamefont{{Bechtol}}},
  \bibinfo{author}{\bibfnamefont{M.}~\bibnamefont{{Ahlers}}},
  \bibinfo{author}{\bibfnamefont{M.}~\bibnamefont{{Di Mauro}}},
  \bibinfo{author}{\bibfnamefont{M.}~\bibnamefont{{Ajello}}}, \bibnamefont{and}
  \bibinfo{author}{\bibfnamefont{J.}~\bibnamefont{{Vandenbroucke}}},
  \bibinfo{journal}{ArXiv e-prints}  (\bibinfo{year}{2015}),
  \eprint{1511.00688}.

\bibitem[{\citenamefont{Anchordoqui
  et~al.}(2014{\natexlab{b}})}]{anchordoqui13}
\bibinfo{author}{\bibfnamefont{L.~A.} \bibnamefont{Anchordoqui}}
  \bibnamefont{et~al.}, \bibinfo{journal}{JHEAp}
  \textbf{\bibinfo{volume}{1-2}}, \bibinfo{pages}{1}
  (\bibinfo{year}{2014}{\natexlab{b}}), \eprint{1312.6587}.

\bibitem[{\citenamefont{{Berezinsky} and {Zatsepin}}(1969)}]{berezinsky}
\bibinfo{author}{\bibfnamefont{V.}~\bibnamefont{{Berezinsky}}}
  \bibnamefont{and}
  \bibinfo{author}{\bibfnamefont{G.}~\bibnamefont{{Zatsepin}}},
  \bibinfo{journal}{Phys. Lett. B} \textbf{\bibinfo{volume}{28}},
  \bibinfo{pages}{423} (\bibinfo{year}{1969}).

\bibitem[{\citenamefont{{Stecker}}(1973)}]{stecker73_cosmogenic}
\bibinfo{author}{\bibfnamefont{F.~W.} \bibnamefont{{Stecker}}},
  \bibinfo{journal}{\apss} \textbf{\bibinfo{volume}{20}}, \bibinfo{pages}{47}
  (\bibinfo{year}{1973}).

\bibitem[{\citenamefont{{Berezinskii} and {Smirnov}}(1975)}]{berezinsky1}
\bibinfo{author}{\bibfnamefont{V.~S.} \bibnamefont{{Berezinskii}}}
  \bibnamefont{and} \bibinfo{author}{\bibfnamefont{A.~I.}
  \bibnamefont{{Smirnov}}}, \bibinfo{journal}{\apss}
  \textbf{\bibinfo{volume}{32}}, \bibinfo{pages}{461} (\bibinfo{year}{1975}).

\bibitem[{\citenamefont{{Engel} et~al.}(2001)\citenamefont{{Engel}, {Seckel},
  and {Stanev}}}]{engel01}
\bibinfo{author}{\bibfnamefont{R.}~\bibnamefont{{Engel}}},
  \bibinfo{author}{\bibfnamefont{D.}~\bibnamefont{{Seckel}}}, \bibnamefont{and}
  \bibinfo{author}{\bibfnamefont{T.}~\bibnamefont{{Stanev}}},
  \bibinfo{journal}{\prd} \textbf{\bibinfo{volume}{64}},
  \bibinfo{pages}{093010} (\bibinfo{year}{2001}), \eprint{astro-ph/0101216}.

\bibitem[{\citenamefont{{Semikoz} and {Sigl}}(2004)}]{semikoz_sigl}
\bibinfo{author}{\bibfnamefont{D.~V.} \bibnamefont{{Semikoz}}}
  \bibnamefont{and} \bibinfo{author}{\bibfnamefont{G.}~\bibnamefont{{Sigl}}},
  \bibinfo{journal}{\jcap} \textbf{\bibinfo{volume}{4}}, \bibinfo{eid}{003}
  (\bibinfo{year}{2004}), \eprint{hep-ph/0309328}.

\bibitem[{\citenamefont{{Allard} et~al.}(2006)\citenamefont{{Allard}, {Ave},
  {Busca}, {Malkan}, {Olinto}, {Parizot}, {Stecker}, and
  {Yamamoto}}}]{allard06}
\bibinfo{author}{\bibfnamefont{D.}~\bibnamefont{{Allard}}},
  \bibinfo{author}{\bibfnamefont{M.}~\bibnamefont{{Ave}}},
  \bibinfo{author}{\bibfnamefont{N.}~\bibnamefont{{Busca}}},
  \bibinfo{author}{\bibfnamefont{M.~A.} \bibnamefont{{Malkan}}},
  \bibinfo{author}{\bibfnamefont{A.~V.} \bibnamefont{{Olinto}}},
  \bibinfo{author}{\bibfnamefont{E.}~\bibnamefont{{Parizot}}},
  \bibinfo{author}{\bibfnamefont{F.~W.} \bibnamefont{{Stecker}}},
  \bibnamefont{and}
  \bibinfo{author}{\bibfnamefont{T.}~\bibnamefont{{Yamamoto}}},
  \bibinfo{journal}{\jcap} \textbf{\bibinfo{volume}{9}}, \bibinfo{eid}{005}
  (\bibinfo{year}{2006}), \eprint{astro-ph/0605327}.

\bibitem[{\citenamefont{{Kotera} et~al.}(2010)\citenamefont{{Kotera}, {Allard},
  and {Olinto}}}]{kotera10}
\bibinfo{author}{\bibfnamefont{K.}~\bibnamefont{{Kotera}}},
  \bibinfo{author}{\bibfnamefont{D.}~\bibnamefont{{Allard}}}, \bibnamefont{and}
  \bibinfo{author}{\bibfnamefont{A.~V.} \bibnamefont{{Olinto}}},
  \bibinfo{journal}{\jcap} \textbf{\bibinfo{volume}{10}}, \bibinfo{eid}{013}
  (\bibinfo{year}{2010}), \eprint{1009.1382}.

\bibitem[{\citenamefont{{Ahlers} et~al.}(2010)\citenamefont{{Ahlers},
  {Anchordoqui}, {Gonzalez-Garcia}, {Halzen}, and {Sarkar}}}]{ahlers10}
\bibinfo{author}{\bibfnamefont{M.}~\bibnamefont{{Ahlers}}},
  \bibinfo{author}{\bibfnamefont{L.~A.} \bibnamefont{{Anchordoqui}}},
  \bibinfo{author}{\bibfnamefont{M.~C.} \bibnamefont{{Gonzalez-Garcia}}},
  \bibinfo{author}{\bibfnamefont{F.}~\bibnamefont{{Halzen}}}, \bibnamefont{and}
  \bibinfo{author}{\bibfnamefont{S.}~\bibnamefont{{Sarkar}}},
  \bibinfo{journal}{Astroparticle Physics} \textbf{\bibinfo{volume}{34}},
  \bibinfo{pages}{106} (\bibinfo{year}{2010}), \eprint{1005.2620}.

\bibitem[{\citenamefont{{Aloisio} et~al.}(2015)\citenamefont{{Aloisio},
  {Boncioli}, {di Matteo}, {Grillo}, {Petrera}, and {Salamida}}}]{aloisio15}
\bibinfo{author}{\bibfnamefont{R.}~\bibnamefont{{Aloisio}}},
  \bibinfo{author}{\bibfnamefont{D.}~\bibnamefont{{Boncioli}}},
  \bibinfo{author}{\bibfnamefont{A.}~\bibnamefont{{di Matteo}}},
  \bibinfo{author}{\bibfnamefont{A.~F.} \bibnamefont{{Grillo}}},
  \bibinfo{author}{\bibfnamefont{S.}~\bibnamefont{{Petrera}}},
  \bibnamefont{and}
  \bibinfo{author}{\bibfnamefont{F.}~\bibnamefont{{Salamida}}},
  \bibinfo{journal}{\jcap} \textbf{\bibinfo{volume}{10}}, \bibinfo{eid}{006}
  (\bibinfo{year}{2015}), \eprint{1505.04020}.

\bibitem[{\citenamefont{{IceCube-Gen2 Collaboration}
  et~al.}(2014)\citenamefont{{IceCube-Gen2 Collaboration}, {:}, {Aartsen},
  {Ackermann}, {Adams}, {Aguilar}, {Ahlers}, {Ahrens}, {Altmann}, {Anderson}
  et~al.}}]{IceCube_gen2}
\bibinfo{author}{\bibnamefont{{IceCube-Gen2 Collaboration}}},
  \bibinfo{author}{\bibnamefont{{:}}}, \bibinfo{author}{\bibfnamefont{M.~G.}
  \bibnamefont{{Aartsen}}},
  \bibinfo{author}{\bibfnamefont{M.}~\bibnamefont{{Ackermann}}},
  \bibinfo{author}{\bibfnamefont{J.}~\bibnamefont{{Adams}}},
  \bibinfo{author}{\bibfnamefont{J.~A.} \bibnamefont{{Aguilar}}},
  \bibinfo{author}{\bibfnamefont{M.}~\bibnamefont{{Ahlers}}},
  \bibinfo{author}{\bibfnamefont{M.}~\bibnamefont{{Ahrens}}},
  \bibinfo{author}{\bibfnamefont{D.}~\bibnamefont{{Altmann}}},
  \bibinfo{author}{\bibfnamefont{T.}~\bibnamefont{{Anderson}}},
  \bibnamefont{et~al.}, \bibinfo{journal}{ArXiv e-prints}
  (\bibinfo{year}{2014}), \eprint{1412.5106}.

\bibitem[{\citenamefont{{Bagley} et~al.}(2009)}]{km3net}
\bibinfo{author}{\bibfnamefont{P.}~\bibnamefont{{Bagley}}} \bibnamefont{et~al.}
  (\bibinfo{year}{2009}),
  \urlprefix\url{http://www.km3net.org/TDR/TDRKM3NeT.pdf}.

\bibitem[{\citenamefont{{Avrorin} et~al.}(2015)\citenamefont{{Avrorin},
  {Avrorin}, {Aynutdinov}, {Bannasch}, {Belolaptikov}, {Bogorodsky},
  {Brudanin}, {Budnev}, {Danilchenko}, {Domogatsky} et~al.}}]{GVD}
\bibinfo{author}{\bibfnamefont{A.~D.} \bibnamefont{{Avrorin}}},
  \bibinfo{author}{\bibfnamefont{A.~V.} \bibnamefont{{Avrorin}}},
  \bibinfo{author}{\bibfnamefont{V.~M.} \bibnamefont{{Aynutdinov}}},
  \bibinfo{author}{\bibfnamefont{R.}~\bibnamefont{{Bannasch}}},
  \bibinfo{author}{\bibfnamefont{I.~A.} \bibnamefont{{Belolaptikov}}},
  \bibinfo{author}{\bibfnamefont{D.~Y.} \bibnamefont{{Bogorodsky}}},
  \bibinfo{author}{\bibfnamefont{V.~B.} \bibnamefont{{Brudanin}}},
  \bibinfo{author}{\bibfnamefont{N.~M.} \bibnamefont{{Budnev}}},
  \bibinfo{author}{\bibfnamefont{I.~A.} \bibnamefont{{Danilchenko}}},
  \bibinfo{author}{\bibfnamefont{G.~V.} \bibnamefont{{Domogatsky}}},
  \bibnamefont{et~al.}, \bibinfo{journal}{Physics of Particles and Nuclei}
  \textbf{\bibinfo{volume}{46}}, \bibinfo{pages}{211} (\bibinfo{year}{2015}).

\bibitem[{\citenamefont{{Kravchenko} et~al.}(2012)\citenamefont{{Kravchenko},
  {Hussain}, {Seckel}, {Besson}, {Fensholt}, {Ralston}, {Taylor}, {Ratzlaff},
  and {Young}}}]{RICE}
\bibinfo{author}{\bibfnamefont{I.}~\bibnamefont{{Kravchenko}}},
  \bibinfo{author}{\bibfnamefont{S.}~\bibnamefont{{Hussain}}},
  \bibinfo{author}{\bibfnamefont{D.}~\bibnamefont{{Seckel}}},
  \bibinfo{author}{\bibfnamefont{D.}~\bibnamefont{{Besson}}},
  \bibinfo{author}{\bibfnamefont{E.}~\bibnamefont{{Fensholt}}},
  \bibinfo{author}{\bibfnamefont{J.}~\bibnamefont{{Ralston}}},
  \bibinfo{author}{\bibfnamefont{J.}~\bibnamefont{{Taylor}}},
  \bibinfo{author}{\bibfnamefont{K.}~\bibnamefont{{Ratzlaff}}},
  \bibnamefont{and} \bibinfo{author}{\bibfnamefont{R.}~\bibnamefont{{Young}}},
  \bibinfo{journal}{\prd} \textbf{\bibinfo{volume}{85}}, \bibinfo{eid}{062004}
  (\bibinfo{year}{2012}), \eprint{1106.1164}.

\bibitem[{\citenamefont{{Ara Collaboration} et~al.}(2012)\citenamefont{{Ara
  Collaboration}, {Allison}, {Auffenberg}, {Bard}, {Beatty}, {Besson},
  {B{\"o}ser}, {Chen}, {Chen}, {Connolly} et~al.}}]{ARA}
\bibinfo{author}{\bibnamefont{{Ara Collaboration}}},
  \bibinfo{author}{\bibfnamefont{P.}~\bibnamefont{{Allison}}},
  \bibinfo{author}{\bibfnamefont{J.}~\bibnamefont{{Auffenberg}}},
  \bibinfo{author}{\bibfnamefont{R.}~\bibnamefont{{Bard}}},
  \bibinfo{author}{\bibfnamefont{J.~J.} \bibnamefont{{Beatty}}},
  \bibinfo{author}{\bibfnamefont{D.~Z.} \bibnamefont{{Besson}}},
  \bibinfo{author}{\bibfnamefont{S.}~\bibnamefont{{B{\"o}ser}}},
  \bibinfo{author}{\bibfnamefont{C.}~\bibnamefont{{Chen}}},
  \bibinfo{author}{\bibfnamefont{P.}~\bibnamefont{{Chen}}},
  \bibinfo{author}{\bibfnamefont{A.}~\bibnamefont{{Connolly}}},
  \bibnamefont{et~al.}, \bibinfo{journal}{Astroparticle Physics}
  \textbf{\bibinfo{volume}{35}}, \bibinfo{pages}{457} (\bibinfo{year}{2012}),
  \eprint{1105.2854}.

\bibitem[{\citenamefont{{Barwick} et~al.}(2015)\citenamefont{{Barwick}, {Berg},
  {Besson}, {Binder}, {Binns}, {Boersma}, {Bose}, {Braun}, {Buckley}, {Bugaev}
  et~al.}}]{ARIANNA}
\bibinfo{author}{\bibfnamefont{S.~W.} \bibnamefont{{Barwick}}},
  \bibinfo{author}{\bibfnamefont{E.~C.} \bibnamefont{{Berg}}},
  \bibinfo{author}{\bibfnamefont{D.~Z.} \bibnamefont{{Besson}}},
  \bibinfo{author}{\bibfnamefont{G.}~\bibnamefont{{Binder}}},
  \bibinfo{author}{\bibfnamefont{W.~R.} \bibnamefont{{Binns}}},
  \bibinfo{author}{\bibfnamefont{D.~J.} \bibnamefont{{Boersma}}},
  \bibinfo{author}{\bibfnamefont{R.~G.} \bibnamefont{{Bose}}},
  \bibinfo{author}{\bibfnamefont{D.~L.} \bibnamefont{{Braun}}},
  \bibinfo{author}{\bibfnamefont{J.~H.} \bibnamefont{{Buckley}}},
  \bibinfo{author}{\bibfnamefont{V.}~\bibnamefont{{Bugaev}}},
  \bibnamefont{et~al.}, \bibinfo{journal}{Astroparticle Physics}
  \textbf{\bibinfo{volume}{70}}, \bibinfo{pages}{12} (\bibinfo{year}{2015}),
  \eprint{1410.7352}.

\bibitem[{\citenamefont{{Fargion}}(2002)}]{fargion02}
\bibinfo{author}{\bibfnamefont{D.}~\bibnamefont{{Fargion}}},
  \bibinfo{journal}{\apj} \textbf{\bibinfo{volume}{570}}, \bibinfo{pages}{909}
  (\bibinfo{year}{2002}), \eprint{astro-ph/0002453}.

\bibitem[{\citenamefont{{Feng} et~al.}(2002)\citenamefont{{Feng}, {Fisher},
  {Wilczek}, and {Yu}}}]{feng02}
\bibinfo{author}{\bibfnamefont{J.~L.} \bibnamefont{{Feng}}},
  \bibinfo{author}{\bibfnamefont{P.}~\bibnamefont{{Fisher}}},
  \bibinfo{author}{\bibfnamefont{F.}~\bibnamefont{{Wilczek}}},
  \bibnamefont{and} \bibinfo{author}{\bibfnamefont{T.~M.} \bibnamefont{{Yu}}},
  \bibinfo{journal}{Physical Review Letters} \textbf{\bibinfo{volume}{88}},
  \bibinfo{eid}{161102} (\bibinfo{year}{2002}), \eprint{hep-ph/0105067}.

\bibitem[{\citenamefont{{Bertou} et~al.}(2002)\citenamefont{{Bertou},
  {Billoir}, {Deligny}, {Lachaud}, and {Letessier-Selvon}}}]{bertou}
\bibinfo{author}{\bibfnamefont{X.}~\bibnamefont{{Bertou}}},
  \bibinfo{author}{\bibfnamefont{P.}~\bibnamefont{{Billoir}}},
  \bibinfo{author}{\bibfnamefont{O.}~\bibnamefont{{Deligny}}},
  \bibinfo{author}{\bibfnamefont{C.}~\bibnamefont{{Lachaud}}},
  \bibnamefont{and}
  \bibinfo{author}{\bibfnamefont{A.}~\bibnamefont{{Letessier-Selvon}}},
  \bibinfo{journal}{Astroparticle Physics} \textbf{\bibinfo{volume}{17}},
  \bibinfo{pages}{183} (\bibinfo{year}{2002}), \eprint{astro-ph/0104452}.

\bibitem[{\citenamefont{{Kusenko} and {Weiler}}(2002)}]{kusenko}
\bibinfo{author}{\bibfnamefont{A.}~\bibnamefont{{Kusenko}}} \bibnamefont{and}
  \bibinfo{author}{\bibfnamefont{T.~J.} \bibnamefont{{Weiler}}},
  \bibinfo{journal}{Physical Review Letters} \textbf{\bibinfo{volume}{88}},
  \bibinfo{eid}{161101} (\bibinfo{year}{2002}), \eprint{hep-ph/0106071}.

\bibitem[{\citenamefont{{Aramo} et~al.}(2005)\citenamefont{{Aramo}, {Insolia},
  {Leonardi}, {Miele}, {Perrone}, {Pisanti}, and {Semikoz}}}]{aramo}
\bibinfo{author}{\bibfnamefont{C.}~\bibnamefont{{Aramo}}},
  \bibinfo{author}{\bibfnamefont{A.}~\bibnamefont{{Insolia}}},
  \bibinfo{author}{\bibfnamefont{A.}~\bibnamefont{{Leonardi}}},
  \bibinfo{author}{\bibfnamefont{G.}~\bibnamefont{{Miele}}},
  \bibinfo{author}{\bibfnamefont{L.}~\bibnamefont{{Perrone}}},
  \bibinfo{author}{\bibfnamefont{O.}~\bibnamefont{{Pisanti}}},
  \bibnamefont{and} \bibinfo{author}{\bibfnamefont{D.~V.}
  \bibnamefont{{Semikoz}}}, \bibinfo{journal}{Astroparticle Physics}
  \textbf{\bibinfo{volume}{23}}, \bibinfo{pages}{65} (\bibinfo{year}{2005}),
  \eprint{astro-ph/0407638}.

\bibitem[{\citenamefont{{Aab} et~al.}(2015)\citenamefont{{Aab}, {Abreu},
  {Aglietta}, {Ahn}, {Al Samarai}, {Albuquerque}, {Allekotte}, {Allison},
  {Almela}, {Alvarez Castillo} et~al.}}]{auger15}
\bibinfo{author}{\bibfnamefont{A.}~\bibnamefont{{Aab}}},
  \bibinfo{author}{\bibfnamefont{P.}~\bibnamefont{{Abreu}}},
  \bibinfo{author}{\bibfnamefont{M.}~\bibnamefont{{Aglietta}}},
  \bibinfo{author}{\bibfnamefont{E.~J.} \bibnamefont{{Ahn}}},
  \bibinfo{author}{\bibfnamefont{I.}~\bibnamefont{{Al Samarai}}},
  \bibinfo{author}{\bibfnamefont{I.~F.~M.} \bibnamefont{{Albuquerque}}},
  \bibinfo{author}{\bibfnamefont{I.}~\bibnamefont{{Allekotte}}},
  \bibinfo{author}{\bibfnamefont{P.}~\bibnamefont{{Allison}}},
  \bibinfo{author}{\bibfnamefont{A.}~\bibnamefont{{Almela}}},
  \bibinfo{author}{\bibfnamefont{J.}~\bibnamefont{{Alvarez Castillo}}},
  \bibnamefont{et~al.}, \bibinfo{journal}{\prd} \textbf{\bibinfo{volume}{91}},
  \bibinfo{eid}{092008} (\bibinfo{year}{2015}), \eprint{1504.05397}.

\bibitem[{\citenamefont{{Gorham} et~al.}(2010)\citenamefont{{Gorham},
  {Allison}, {Baughman}, {Beatty}, {Belov}, {Besson}, {Bevan}, {Binns}, {Chen},
  {Chen} et~al.}}]{ANITA}
\bibinfo{author}{\bibfnamefont{P.~W.} \bibnamefont{{Gorham}}},
  \bibinfo{author}{\bibfnamefont{P.}~\bibnamefont{{Allison}}},
  \bibinfo{author}{\bibfnamefont{B.~M.} \bibnamefont{{Baughman}}},
  \bibinfo{author}{\bibfnamefont{J.~J.} \bibnamefont{{Beatty}}},
  \bibinfo{author}{\bibfnamefont{K.}~\bibnamefont{{Belov}}},
  \bibinfo{author}{\bibfnamefont{D.~Z.} \bibnamefont{{Besson}}},
  \bibinfo{author}{\bibfnamefont{S.}~\bibnamefont{{Bevan}}},
  \bibinfo{author}{\bibfnamefont{W.~R.} \bibnamefont{{Binns}}},
  \bibinfo{author}{\bibfnamefont{C.}~\bibnamefont{{Chen}}},
  \bibinfo{author}{\bibfnamefont{P.}~\bibnamefont{{Chen}}},
  \bibnamefont{et~al.}, \bibinfo{journal}{\prd} \textbf{\bibinfo{volume}{82}},
  \bibinfo{eid}{022004} (\bibinfo{year}{2010}), \eprint{1003.2961}.

\bibitem[{\citenamefont{{Gorham} et~al.}(2011)\citenamefont{{Gorham},
  {Baginski}, {Allison}, {Liewer}, {Miki}, {Hill}, and {Varner}}}]{EVA}
\bibinfo{author}{\bibfnamefont{P.~W.} \bibnamefont{{Gorham}}},
  \bibinfo{author}{\bibfnamefont{F.~E.} \bibnamefont{{Baginski}}},
  \bibinfo{author}{\bibfnamefont{P.}~\bibnamefont{{Allison}}},
  \bibinfo{author}{\bibfnamefont{K.~M.} \bibnamefont{{Liewer}}},
  \bibinfo{author}{\bibfnamefont{C.}~\bibnamefont{{Miki}}},
  \bibinfo{author}{\bibfnamefont{B.}~\bibnamefont{{Hill}}}, \bibnamefont{and}
  \bibinfo{author}{\bibfnamefont{G.~S.} \bibnamefont{{Varner}}},
  \bibinfo{journal}{Astroparticle Physics} \textbf{\bibinfo{volume}{35}},
  \bibinfo{pages}{242} (\bibinfo{year}{2011}), \eprint{1102.3883}.

\bibitem[{\citenamefont{{Martineau-Huynh}
  et~al.}(2016)\citenamefont{{Martineau-Huynh}, {Kotera}, {Bustamente},
  {Charrier}, {De Jong}, {de Vries}, {Fang}, {Feng}, {Finley}, {Gou}
  et~al.}}]{grand}
\bibinfo{author}{\bibfnamefont{O.}~\bibnamefont{{Martineau-Huynh}}},
  \bibinfo{author}{\bibfnamefont{K.}~\bibnamefont{{Kotera}}},
  \bibinfo{author}{\bibfnamefont{M.}~\bibnamefont{{Bustamente}}},
  \bibinfo{author}{\bibfnamefont{D.}~\bibnamefont{{Charrier}}},
  \bibinfo{author}{\bibfnamefont{S.}~\bibnamefont{{De Jong}}},
  \bibinfo{author}{\bibfnamefont{K.~D.} \bibnamefont{{de Vries}}},
  \bibinfo{author}{\bibfnamefont{K.}~\bibnamefont{{Fang}}},
  \bibinfo{author}{\bibfnamefont{Z.}~\bibnamefont{{Feng}}},
  \bibinfo{author}{\bibfnamefont{C.}~\bibnamefont{{Finley}}},
  \bibinfo{author}{\bibfnamefont{Q.}~\bibnamefont{{Gou}}},
  \bibnamefont{et~al.}, in \emph{\bibinfo{booktitle}{European Physical Journal
  Web of Conferences}} (\bibinfo{year}{2016}), vol. \bibinfo{volume}{116} of
  \emph{\bibinfo{series}{European Physical Journal Web of Conferences}}, p.
  \bibinfo{pages}{03005}, \eprint{1508.01919}.

\bibitem[{\citenamefont{{Asaoka} and {Sasaki}}(2013)}]{ashra}
\bibinfo{author}{\bibfnamefont{Y.}~\bibnamefont{{Asaoka}}} \bibnamefont{and}
  \bibinfo{author}{\bibfnamefont{M.}~\bibnamefont{{Sasaki}}},
  \bibinfo{journal}{Astroparticle Physics} \textbf{\bibinfo{volume}{41}},
  \bibinfo{pages}{7} (\bibinfo{year}{2013}), \eprint{1202.5656}.

\bibitem[{\citenamefont{{G{\'o}ra} et~al.}(2015)\citenamefont{{G{\'o}ra},
  {Bernardini}, and {Kappes}}}]{gora15}
\bibinfo{author}{\bibfnamefont{D.}~\bibnamefont{{G{\'o}ra}}},
  \bibinfo{author}{\bibfnamefont{E.}~\bibnamefont{{Bernardini}}},
  \bibnamefont{and} \bibinfo{author}{\bibfnamefont{A.}~\bibnamefont{{Kappes}}},
  \bibinfo{journal}{Astroparticle Physics} \textbf{\bibinfo{volume}{61}},
  \bibinfo{pages}{12} (\bibinfo{year}{2015}), \eprint{1402.4243}.

\bibitem[{\citenamefont{Gora and Bernardini}(2016)}]{gora16}
\bibinfo{author}{\bibfnamefont{D.}~\bibnamefont{Gora}} \bibnamefont{and}
  \bibinfo{author}{\bibfnamefont{E.}~\bibnamefont{Bernardini}}
  (\bibinfo{year}{2016}), \eprint{1606.01676}.

\bibitem[{\citenamefont{{Gonzalez-Garcia}
  et~al.}(2014)\citenamefont{{Gonzalez-Garcia}, {Maltoni}, and
  {Schwetz}}}]{Gonzalez-Garcia:2014bfa}
\bibinfo{author}{\bibfnamefont{M.~C.} \bibnamefont{{Gonzalez-Garcia}}},
  \bibinfo{author}{\bibfnamefont{M.}~\bibnamefont{{Maltoni}}},
  \bibnamefont{and}
  \bibinfo{author}{\bibfnamefont{T.}~\bibnamefont{{Schwetz}}},
  \bibinfo{journal}{Journal of High Energy Physics}
  \textbf{\bibinfo{volume}{11}}, \bibinfo{eid}{52} (\bibinfo{year}{2014}),
  \eprint{1409.5439}.

\bibitem[{\citenamefont{{Anchordoqui} et~al.}(2004)\citenamefont{{Anchordoqui},
  {Goldberg}, {Halzen}, and {Weiler}}}]{Anchordoqui:2003vc}
\bibinfo{author}{\bibfnamefont{L.~A.} \bibnamefont{{Anchordoqui}}},
  \bibinfo{author}{\bibfnamefont{H.}~\bibnamefont{{Goldberg}}},
  \bibinfo{author}{\bibfnamefont{F.}~\bibnamefont{{Halzen}}}, \bibnamefont{and}
  \bibinfo{author}{\bibfnamefont{T.~J.} \bibnamefont{{Weiler}}},
  \bibinfo{journal}{Physics Letters B} \textbf{\bibinfo{volume}{593}},
  \bibinfo{pages}{42} (\bibinfo{year}{2004}), \eprint{astro-ph/0311002}.

\bibitem[{\citenamefont{{Aartsen}
  et~al.}(2015{\natexlab{c}})\citenamefont{{Aartsen}, {Ackermann}, {Adams},
  {Aguilar}, {Ahlers}, {Ahrens}, {Altmann}, {Anderson}, {Arguelles}, {Arlen}
  et~al.}}]{Aartsen:2015ivb}
\bibinfo{author}{\bibfnamefont{M.~G.} \bibnamefont{{Aartsen}}},
  \bibinfo{author}{\bibfnamefont{M.}~\bibnamefont{{Ackermann}}},
  \bibinfo{author}{\bibfnamefont{J.}~\bibnamefont{{Adams}}},
  \bibinfo{author}{\bibfnamefont{J.~A.} \bibnamefont{{Aguilar}}},
  \bibinfo{author}{\bibfnamefont{M.}~\bibnamefont{{Ahlers}}},
  \bibinfo{author}{\bibfnamefont{M.}~\bibnamefont{{Ahrens}}},
  \bibinfo{author}{\bibfnamefont{D.}~\bibnamefont{{Altmann}}},
  \bibinfo{author}{\bibfnamefont{T.}~\bibnamefont{{Anderson}}},
  \bibinfo{author}{\bibfnamefont{C.}~\bibnamefont{{Arguelles}}},
  \bibinfo{author}{\bibfnamefont{T.~C.} \bibnamefont{{Arlen}}},
  \bibnamefont{et~al.}, \bibinfo{journal}{Physical Review Letters}
  \textbf{\bibinfo{volume}{114}}, \bibinfo{eid}{171102}
  (\bibinfo{year}{2015}{\natexlab{c}}), \eprint{1502.03376}.

\bibitem[{\citenamefont{{Armesto} et~al.}(2008)\citenamefont{{Armesto},
  {Merino}, {Parente}, and {Zas}}}]{cross_section_loss}
\bibinfo{author}{\bibfnamefont{N.}~\bibnamefont{{Armesto}}},
  \bibinfo{author}{\bibfnamefont{C.}~\bibnamefont{{Merino}}},
  \bibinfo{author}{\bibfnamefont{G.}~\bibnamefont{{Parente}}},
  \bibnamefont{and} \bibinfo{author}{\bibfnamefont{E.}~\bibnamefont{{Zas}}},
  \bibinfo{journal}{\prd} \textbf{\bibinfo{volume}{77}}, \bibinfo{eid}{013001}
  (\bibinfo{year}{2008}), \eprint{0709.4461}.

\bibitem[{\citenamefont{{Gandhi} et~al.}(1996)\citenamefont{{Gandhi}, {Quigg},
  {Hall Reno}, and {Sarcevic}}}]{Gandhi:1995tf}
\bibinfo{author}{\bibfnamefont{R.}~\bibnamefont{{Gandhi}}},
  \bibinfo{author}{\bibfnamefont{C.}~\bibnamefont{{Quigg}}},
  \bibinfo{author}{\bibfnamefont{M.}~\bibnamefont{{Hall Reno}}},
  \bibnamefont{and}
  \bibinfo{author}{\bibfnamefont{I.}~\bibnamefont{{Sarcevic}}},
  \bibinfo{journal}{Astroparticle Physics} \textbf{\bibinfo{volume}{5}},
  \bibinfo{pages}{81} (\bibinfo{year}{1996}), \eprint{hep-ph/9512364}.

\bibitem[{\citenamefont{Cooper-Sarkar and Sarkar}(2008)}]{CooperSarkar:2007cv}
\bibinfo{author}{\bibfnamefont{A.}~\bibnamefont{Cooper-Sarkar}}
  \bibnamefont{and} \bibinfo{author}{\bibfnamefont{S.}~\bibnamefont{Sarkar}},
  \bibinfo{journal}{JHEP} \textbf{\bibinfo{volume}{01}}, \bibinfo{pages}{075}
  (\bibinfo{year}{2008}), \eprint{0710.5303}.

\bibitem[{\citenamefont{Connolly et~al.}(2011)\citenamefont{Connolly, Thorne,
  and Waters}}]{Connolly:2011vc}
\bibinfo{author}{\bibfnamefont{A.}~\bibnamefont{Connolly}},
  \bibinfo{author}{\bibfnamefont{R.~S.} \bibnamefont{Thorne}},
  \bibnamefont{and} \bibinfo{author}{\bibfnamefont{D.}~\bibnamefont{Waters}},
  \bibinfo{journal}{Phys. Rev.} \textbf{\bibinfo{volume}{D83}},
  \bibinfo{pages}{113009} (\bibinfo{year}{2011}), \eprint{1102.0691}.

\bibitem[{\citenamefont{Argüelles et~al.}(2015)\citenamefont{Argüelles, Halzen,
  Wille, Kroll, and Reno}}]{Arguelles:2015wba}
\bibinfo{author}{\bibfnamefont{C.~A.} \bibnamefont{Argüelles}},
  \bibinfo{author}{\bibfnamefont{F.}~\bibnamefont{Halzen}},
  \bibinfo{author}{\bibfnamefont{L.}~\bibnamefont{Wille}},
  \bibinfo{author}{\bibfnamefont{M.}~\bibnamefont{Kroll}}, \bibnamefont{and}
  \bibinfo{author}{\bibfnamefont{M.~H.} \bibnamefont{Reno}},
  \bibinfo{journal}{Phys. Rev.} \textbf{\bibinfo{volume}{D92}},
  \bibinfo{pages}{074040} (\bibinfo{year}{2015}), \eprint{1504.06639}.

\bibitem[{\citenamefont{Cooper-Sarkar et~al.}(2011)\citenamefont{Cooper-Sarkar,
  Mertsch, and Sarkar}}]{CooperSarkar:2011pa}
\bibinfo{author}{\bibfnamefont{A.}~\bibnamefont{Cooper-Sarkar}},
  \bibinfo{author}{\bibfnamefont{P.}~\bibnamefont{Mertsch}}, \bibnamefont{and}
  \bibinfo{author}{\bibfnamefont{S.}~\bibnamefont{Sarkar}},
  \bibinfo{journal}{JHEP} \textbf{\bibinfo{volume}{08}}, \bibinfo{pages}{042}
  (\bibinfo{year}{2011}), \eprint{1106.3723}.

\bibitem[{\citenamefont{Fargion}(2002)}]{Fargion:2000iz}
\bibinfo{author}{\bibfnamefont{D.}~\bibnamefont{Fargion}},
  \bibinfo{journal}{Astrophys. J.} \textbf{\bibinfo{volume}{570}},
  \bibinfo{pages}{909} (\bibinfo{year}{2002}), \eprint{astro-ph/0002453}.

\bibitem[{\citenamefont{Aramo et~al.}(2005)\citenamefont{Aramo, Insolia,
  Leonardi, Miele, Perrone, Pisanti, and Semikoz}}]{Aramo:2004pr}
\bibinfo{author}{\bibfnamefont{C.}~\bibnamefont{Aramo}},
  \bibinfo{author}{\bibfnamefont{A.}~\bibnamefont{Insolia}},
  \bibinfo{author}{\bibfnamefont{A.}~\bibnamefont{Leonardi}},
  \bibinfo{author}{\bibfnamefont{G.}~\bibnamefont{Miele}},
  \bibinfo{author}{\bibfnamefont{L.}~\bibnamefont{Perrone}},
  \bibinfo{author}{\bibfnamefont{O.}~\bibnamefont{Pisanti}}, \bibnamefont{and}
  \bibinfo{author}{\bibfnamefont{D.~V.} \bibnamefont{Semikoz}},
  \bibinfo{journal}{Astropart. Phys.} \textbf{\bibinfo{volume}{23}},
  \bibinfo{pages}{65} (\bibinfo{year}{2005}), \eprint{astro-ph/0407638}.

\bibitem[{\citenamefont{Dutta et~al.}(2005)\citenamefont{Dutta, Huang, and
  Reno}}]{Dutta:2005yt}
\bibinfo{author}{\bibfnamefont{S.~I.} \bibnamefont{Dutta}},
  \bibinfo{author}{\bibfnamefont{Y.}~\bibnamefont{Huang}}, \bibnamefont{and}
  \bibinfo{author}{\bibfnamefont{M.~H.} \bibnamefont{Reno}},
  \bibinfo{journal}{Phys. Rev.} \textbf{\bibinfo{volume}{D72}},
  \bibinfo{pages}{013005} (\bibinfo{year}{2005}), \eprint{hep-ph/0504208}.

\bibitem[{\citenamefont{Bigas et~al.}(2008)\citenamefont{Bigas, Deligny, Payet,
  and Van~Elewyck}}]{Bigas:2008ff}
\bibinfo{author}{\bibfnamefont{O.~B.} \bibnamefont{Bigas}},
  \bibinfo{author}{\bibfnamefont{O.}~\bibnamefont{Deligny}},
  \bibinfo{author}{\bibfnamefont{K.}~\bibnamefont{Payet}}, \bibnamefont{and}
  \bibinfo{author}{\bibfnamefont{V.}~\bibnamefont{Van~Elewyck}},
  \bibinfo{journal}{Phys. Rev.} \textbf{\bibinfo{volume}{D77}},
  \bibinfo{pages}{103004} (\bibinfo{year}{2008}), \eprint{0802.1119}.

\bibitem[{\citenamefont{{Olive}}(2014)}]{pdg}
\bibinfo{author}{\bibfnamefont{K.~e.~a.} \bibnamefont{{Olive}}},
  \bibinfo{journal}{Chin. Phys. C} \textbf{\bibinfo{volume}{38}},
  \bibinfo{pages}{090001} (\bibinfo{year}{2014}).

\bibitem[{\citenamefont{{Adams}
  et~al.}(2015{\natexlab{a}})\citenamefont{{Adams}, {Ahmad}, {Albert},
  {Allard}, {Anchordoqui}, {Andreev}, {Anzalone}, {Arai}, {Asano}, {Ave Pernas}
  et~al.}}]{jemeuso}
\bibinfo{author}{\bibfnamefont{J.~H.} \bibnamefont{{Adams}}},
  \bibinfo{author}{\bibfnamefont{S.}~\bibnamefont{{Ahmad}}},
  \bibinfo{author}{\bibfnamefont{J.-N.} \bibnamefont{{Albert}}},
  \bibinfo{author}{\bibfnamefont{D.}~\bibnamefont{{Allard}}},
  \bibinfo{author}{\bibfnamefont{L.}~\bibnamefont{{Anchordoqui}}},
  \bibinfo{author}{\bibfnamefont{V.}~\bibnamefont{{Andreev}}},
  \bibinfo{author}{\bibfnamefont{A.}~\bibnamefont{{Anzalone}}},
  \bibinfo{author}{\bibfnamefont{Y.}~\bibnamefont{{Arai}}},
  \bibinfo{author}{\bibfnamefont{K.}~\bibnamefont{{Asano}}},
  \bibinfo{author}{\bibfnamefont{M.}~\bibnamefont{{Ave Pernas}}},
  \bibnamefont{et~al.}, \bibinfo{journal}{Experimental Astronomy}
  \textbf{\bibinfo{volume}{40}}, \bibinfo{pages}{19}
  (\bibinfo{year}{2015}{\natexlab{a}}).

\bibitem[{\citenamefont{{Stecker} et~al.}(2004)\citenamefont{{Stecker},
  {Krizmanic}, {Barbier}, {Loh}, {Mitchell}, {Sokolsky}, and
  {Streitmatter}}}]{OWL}
\bibinfo{author}{\bibfnamefont{F.~W.} \bibnamefont{{Stecker}}},
  \bibinfo{author}{\bibfnamefont{J.~F.} \bibnamefont{{Krizmanic}}},
  \bibinfo{author}{\bibfnamefont{L.~M.} \bibnamefont{{Barbier}}},
  \bibinfo{author}{\bibfnamefont{E.}~\bibnamefont{{Loh}}},
  \bibinfo{author}{\bibfnamefont{J.~W.} \bibnamefont{{Mitchell}}},
  \bibinfo{author}{\bibfnamefont{P.}~\bibnamefont{{Sokolsky}}},
  \bibnamefont{and} \bibinfo{author}{\bibfnamefont{R.~E.}
  \bibnamefont{{Streitmatter}}}, \bibinfo{journal}{Nuclear Physics B
  Proceedings Supplements} \textbf{\bibinfo{volume}{136}}, \bibinfo{pages}{433}
  (\bibinfo{year}{2004}), \eprint{astro-ph/0408162}.

\bibitem[{\citenamefont{{Adams}
  et~al.}(2015{\natexlab{b}})\citenamefont{{Adams}, {Ahmad}, {Albert},
  {Allard}, {Anchordoqui}, {Andreev}, {Anzalone}, {Arai}, {Asano}, {Ave Pernas}
  et~al.}}]{euso_ams}
\bibinfo{author}{\bibfnamefont{J.~H.} \bibnamefont{{Adams}}},
  \bibinfo{author}{\bibfnamefont{S.}~\bibnamefont{{Ahmad}}},
  \bibinfo{author}{\bibfnamefont{J.-N.} \bibnamefont{{Albert}}},
  \bibinfo{author}{\bibfnamefont{D.}~\bibnamefont{{Allard}}},
  \bibinfo{author}{\bibfnamefont{L.}~\bibnamefont{{Anchordoqui}}},
  \bibinfo{author}{\bibfnamefont{V.}~\bibnamefont{{Andreev}}},
  \bibinfo{author}{\bibfnamefont{A.}~\bibnamefont{{Anzalone}}},
  \bibinfo{author}{\bibfnamefont{Y.}~\bibnamefont{{Arai}}},
  \bibinfo{author}{\bibfnamefont{K.}~\bibnamefont{{Asano}}},
  \bibinfo{author}{\bibfnamefont{M.}~\bibnamefont{{Ave Pernas}}},
  \bibnamefont{et~al.}, \bibinfo{journal}{Experimental Astronomy}
  \textbf{\bibinfo{volume}{40}}, \bibinfo{pages}{45}
  (\bibinfo{year}{2015}{\natexlab{b}}), \eprint{1402.6097}.

\bibitem[{\citenamefont{{de la Calle Perez} and
  {Biller}}(2006)}]{wide_field_telescope}
\bibinfo{author}{\bibfnamefont{I.}~\bibnamefont{{de la Calle Perez}}}
  \bibnamefont{and} \bibinfo{author}{\bibfnamefont{S.}~\bibnamefont{{Biller}}},
  \bibinfo{journal}{Astroparticle Physics} \textbf{\bibinfo{volume}{26}},
  \bibinfo{pages}{69} (\bibinfo{year}{2006}), \eprint{astro-ph/0506729}.

\bibitem[{\citenamefont{{Nerling} et~al.}(2006)\citenamefont{{Nerling},
  {Bl{\"u}mer}, {Engel}, and {Risse}}}]{cherenkov_lateral_profile}
\bibinfo{author}{\bibfnamefont{F.}~\bibnamefont{{Nerling}}},
  \bibinfo{author}{\bibfnamefont{J.}~\bibnamefont{{Bl{\"u}mer}}},
  \bibinfo{author}{\bibfnamefont{R.}~\bibnamefont{{Engel}}}, \bibnamefont{and}
  \bibinfo{author}{\bibfnamefont{M.}~\bibnamefont{{Risse}}},
  \bibinfo{journal}{Astroparticle Physics} \textbf{\bibinfo{volume}{24}},
  \bibinfo{pages}{421} (\bibinfo{year}{2006}), \eprint{astro-ph/0506729}.

\bibitem[{\citenamefont{{Winkler} et~al.}(2013)\citenamefont{{Winkler},
  {Tackett}, {Getzewich}, {Liu}, and {Vaughan}}}]{aerosol}
\bibinfo{author}{\bibfnamefont{D.}~\bibnamefont{{Winkler}}},
  \bibinfo{author}{\bibfnamefont{J.}~\bibnamefont{{Tackett}}},
  \bibinfo{author}{\bibfnamefont{B.}~\bibnamefont{{Getzewich}}},
  \bibinfo{author}{\bibfnamefont{Z.}~\bibnamefont{{Liu}}}, \bibnamefont{and}
  \bibinfo{author}{\bibfnamefont{R.}~\bibnamefont{{Vaughan}},
  \bibfnamefont{M.A.~{Rogers}}}, \bibinfo{journal}{Atmos. Chem. Phys.}
  \textbf{\bibinfo{volume}{13}}, \bibinfo{pages}{3345} (\bibinfo{year}{2013}).

\bibitem[{\citenamefont{{Louedec} and {Urban}}(2012)}]{aerosol_phase_function}
\bibinfo{author}{\bibfnamefont{K.}~\bibnamefont{{Louedec}}} \bibnamefont{and}
  \bibinfo{author}{\bibfnamefont{M.}~\bibnamefont{{Urban}}},
  \bibinfo{journal}{\ao} \textbf{\bibinfo{volume}{51}}, \bibinfo{pages}{7842}
  (\bibinfo{year}{2012}), \eprint{1211.0878}.

\bibitem[{\citenamefont{{Tchernin}
  et~al.}(2013{\natexlab{b}})\citenamefont{{Tchernin}, {Aguilar}, {Neronov},
  and {Montaruli}}}]{aguilar13}
\bibinfo{author}{\bibfnamefont{C.}~\bibnamefont{{Tchernin}}},
  \bibinfo{author}{\bibfnamefont{J.~A.} \bibnamefont{{Aguilar}}},
  \bibinfo{author}{\bibfnamefont{A.}~\bibnamefont{{Neronov}}},
  \bibnamefont{and}
  \bibinfo{author}{\bibfnamefont{T.}~\bibnamefont{{Montaruli}}},
  \bibinfo{journal}{\aap} \textbf{\bibinfo{volume}{555}}, \bibinfo{eid}{A70}
  (\bibinfo{year}{2013}{\natexlab{b}}), \eprint{1305.3524}.

\bibitem[{\citenamefont{{Hillas}}(2013)}]{hillas}
\bibinfo{author}{\bibfnamefont{A.~M.} \bibnamefont{{Hillas}}},
  \bibinfo{journal}{Astroparticle Physics} \textbf{\bibinfo{volume}{43}},
  \bibinfo{pages}{19} (\bibinfo{year}{2013}).

\bibitem[{\citenamefont{{Adams} et~al.}(2013)\citenamefont{{Adams}, {Ahmad},
  {Albert}, {Allard}, {Ambrosio}, {Anchordoqui}, {Anzalone}, {Arai}, {Aramo},
  {Asano} et~al.}}]{euso_astropart}
\bibinfo{author}{\bibfnamefont{J.~H.} \bibnamefont{{Adams}}},
  \bibinfo{author}{\bibfnamefont{S.}~\bibnamefont{{Ahmad}}},
  \bibinfo{author}{\bibfnamefont{J.-N.} \bibnamefont{{Albert}}},
  \bibinfo{author}{\bibfnamefont{D.}~\bibnamefont{{Allard}}},
  \bibinfo{author}{\bibfnamefont{M.}~\bibnamefont{{Ambrosio}}},
  \bibinfo{author}{\bibfnamefont{L.}~\bibnamefont{{Anchordoqui}}},
  \bibinfo{author}{\bibfnamefont{A.}~\bibnamefont{{Anzalone}}},
  \bibinfo{author}{\bibfnamefont{Y.}~\bibnamefont{{Arai}}},
  \bibinfo{author}{\bibfnamefont{C.}~\bibnamefont{{Aramo}}},
  \bibinfo{author}{\bibfnamefont{K.}~\bibnamefont{{Asano}}},
  \bibnamefont{et~al.}, \bibinfo{journal}{Astroparticle Physics}
  \textbf{\bibinfo{volume}{44}}, \bibinfo{pages}{76} (\bibinfo{year}{2013}),
  \eprint{1305.2478}.

\bibitem[{\citenamefont{{Adams}
  et~al.}(2015{\natexlab{c}})\citenamefont{{Adams}, {Ahmad}, {Albert},
  {Allard}, {Anchordoqui}, {Andreev}, {Anzalone}, {Arai}, {Asano}, {Ave Pernas}
  et~al.}}]{jemeuso_cloudy}
\bibinfo{author}{\bibfnamefont{J.~H.} \bibnamefont{{Adams}}},
  \bibinfo{author}{\bibfnamefont{S.}~\bibnamefont{{Ahmad}}},
  \bibinfo{author}{\bibfnamefont{J.-N.} \bibnamefont{{Albert}}},
  \bibinfo{author}{\bibfnamefont{D.}~\bibnamefont{{Allard}}},
  \bibinfo{author}{\bibfnamefont{L.}~\bibnamefont{{Anchordoqui}}},
  \bibinfo{author}{\bibfnamefont{V.}~\bibnamefont{{Andreev}}},
  \bibinfo{author}{\bibfnamefont{A.}~\bibnamefont{{Anzalone}}},
  \bibinfo{author}{\bibfnamefont{Y.}~\bibnamefont{{Arai}}},
  \bibinfo{author}{\bibfnamefont{K.}~\bibnamefont{{Asano}}},
  \bibinfo{author}{\bibfnamefont{M.}~\bibnamefont{{Ave Pernas}}},
  \bibnamefont{et~al.}, \bibinfo{journal}{Experimental Astronomy}
  \textbf{\bibinfo{volume}{40}}, \bibinfo{pages}{135}
  (\bibinfo{year}{2015}{\natexlab{c}}).

\bibitem[{\citenamefont{{Broadfoot} and {Kendall}}(1968)}]{airglow}
\bibinfo{author}{\bibfnamefont{A.~L.} \bibnamefont{{Broadfoot}}}
  \bibnamefont{and} \bibinfo{author}{\bibfnamefont{K.~R.}
  \bibnamefont{{Kendall}}}, \bibinfo{journal}{\jgr}
  \textbf{\bibinfo{volume}{73}}, \bibinfo{pages}{426} (\bibinfo{year}{1968}).

\bibitem[{\citenamefont{{Adams}
  et~al.}(2015{\natexlab{d}})\citenamefont{{Adams}, {Ahmad}, {Albert},
  {Allard}, {Anchordoqui}, {Andreev}, {Anzalone}, {Arai}, {Asano}, {Ave Pernas}
  et~al.}}]{TUS}
\bibinfo{author}{\bibfnamefont{J.~H.} \bibnamefont{{Adams}}},
  \bibinfo{author}{\bibfnamefont{S.}~\bibnamefont{{Ahmad}}},
  \bibinfo{author}{\bibfnamefont{J.-N.} \bibnamefont{{Albert}}},
  \bibinfo{author}{\bibfnamefont{D.}~\bibnamefont{{Allard}}},
  \bibinfo{author}{\bibfnamefont{L.}~\bibnamefont{{Anchordoqui}}},
  \bibinfo{author}{\bibfnamefont{V.}~\bibnamefont{{Andreev}}},
  \bibinfo{author}{\bibfnamefont{A.}~\bibnamefont{{Anzalone}}},
  \bibinfo{author}{\bibfnamefont{Y.}~\bibnamefont{{Arai}}},
  \bibinfo{author}{\bibfnamefont{K.}~\bibnamefont{{Asano}}},
  \bibinfo{author}{\bibfnamefont{M.}~\bibnamefont{{Ave Pernas}}},
  \bibnamefont{et~al.}, \bibinfo{journal}{Experimental Astronomy}
  \textbf{\bibinfo{volume}{40}}, \bibinfo{pages}{315}
  (\bibinfo{year}{2015}{\natexlab{d}}).

\bibitem[{\citenamefont{{Leinert} et~al.}(1998)\citenamefont{{Leinert},
  {Bowyer}, {Haikala}, {Hanner}, {Hauser}, {Levasseur-Regourd}, {Mann},
  {Mattila}, {Reach}, {Schlosser} et~al.}}]{airglow_ref}
\bibinfo{author}{\bibfnamefont{C.}~\bibnamefont{{Leinert}}},
  \bibinfo{author}{\bibfnamefont{S.}~\bibnamefont{{Bowyer}}},
  \bibinfo{author}{\bibfnamefont{L.~K.} \bibnamefont{{Haikala}}},
  \bibinfo{author}{\bibfnamefont{M.~S.} \bibnamefont{{Hanner}}},
  \bibinfo{author}{\bibfnamefont{M.~G.} \bibnamefont{{Hauser}}},
  \bibinfo{author}{\bibfnamefont{A.-C.} \bibnamefont{{Levasseur-Regourd}}},
  \bibinfo{author}{\bibfnamefont{I.}~\bibnamefont{{Mann}}},
  \bibinfo{author}{\bibfnamefont{K.}~\bibnamefont{{Mattila}}},
  \bibinfo{author}{\bibfnamefont{W.~T.} \bibnamefont{{Reach}}},
  \bibinfo{author}{\bibfnamefont{W.}~\bibnamefont{{Schlosser}}},
  \bibnamefont{et~al.}, \bibinfo{journal}{\aaps}
  \textbf{\bibinfo{volume}{127}}, \bibinfo{pages}{1} (\bibinfo{year}{1998}).

\bibitem[{\citenamefont{{Cinzano} et~al.}(2001)\citenamefont{{Cinzano},
  {Falchi}, and {Elvidge}}}]{citylights}
\bibinfo{author}{\bibfnamefont{P.}~\bibnamefont{{Cinzano}}},
  \bibinfo{author}{\bibfnamefont{F.}~\bibnamefont{{Falchi}}}, \bibnamefont{and}
  \bibinfo{author}{\bibfnamefont{C.~D.} \bibnamefont{{Elvidge}}},
  \bibinfo{journal}{\mnras} \textbf{\bibinfo{volume}{328}},
  \bibinfo{pages}{689} (\bibinfo{year}{2001}), \eprint{astro-ph/0108052}.

\bibitem[{\citenamefont{{Adams}
  et~al.}(2015{\natexlab{e}})\citenamefont{{Adams}, {Ahmad}, {Albert},
  {Allard}, {Anchordoqui}, {Andreev}, {Anzalone}, {Arai}, {Asano}, {Ave Pernas}
  et~al.}}]{euso_slow_mode}
\bibinfo{author}{\bibfnamefont{J.~H.} \bibnamefont{{Adams}}},
  \bibinfo{author}{\bibfnamefont{S.}~\bibnamefont{{Ahmad}}},
  \bibinfo{author}{\bibfnamefont{J.-N.} \bibnamefont{{Albert}}},
  \bibinfo{author}{\bibfnamefont{D.}~\bibnamefont{{Allard}}},
  \bibinfo{author}{\bibfnamefont{L.}~\bibnamefont{{Anchordoqui}}},
  \bibinfo{author}{\bibfnamefont{V.}~\bibnamefont{{Andreev}}},
  \bibinfo{author}{\bibfnamefont{A.}~\bibnamefont{{Anzalone}}},
  \bibinfo{author}{\bibfnamefont{Y.}~\bibnamefont{{Arai}}},
  \bibinfo{author}{\bibfnamefont{K.}~\bibnamefont{{Asano}}},
  \bibinfo{author}{\bibfnamefont{M.}~\bibnamefont{{Ave Pernas}}},
  \bibnamefont{et~al.}, \bibinfo{journal}{Experimental Astronomy}
  \textbf{\bibinfo{volume}{40}}, \bibinfo{pages}{45}
  (\bibinfo{year}{2015}{\natexlab{e}}), \eprint{1402.6097}.

\bibitem[{\citenamefont{{Knoetig} et~al.}(2013)\citenamefont{{Knoetig},
  {Biland}, {Bretz}, {Bu{\ss}}, {Dorner}, {Einecke}, {Eisenacher},
  {Hildebrand}, {Kr{\"a}henb{\"u}hl}, {Lustermann} et~al.}}]{fact1}
\bibinfo{author}{\bibfnamefont{M.~L.} \bibnamefont{{Knoetig}}},
  \bibinfo{author}{\bibfnamefont{A.}~\bibnamefont{{Biland}}},
  \bibinfo{author}{\bibfnamefont{T.}~\bibnamefont{{Bretz}}},
  \bibinfo{author}{\bibfnamefont{J.}~\bibnamefont{{Bu{\ss}}}},
  \bibinfo{author}{\bibfnamefont{D.}~\bibnamefont{{Dorner}}},
  \bibinfo{author}{\bibfnamefont{S.}~\bibnamefont{{Einecke}}},
  \bibinfo{author}{\bibfnamefont{D.}~\bibnamefont{{Eisenacher}}},
  \bibinfo{author}{\bibfnamefont{D.}~\bibnamefont{{Hildebrand}}},
  \bibinfo{author}{\bibfnamefont{T.}~\bibnamefont{{Kr{\"a}henb{\"u}hl}}},
  \bibinfo{author}{\bibfnamefont{W.}~\bibnamefont{{Lustermann}}},
  \bibnamefont{et~al.}, \bibinfo{journal}{ArXiv e-prints}
  (\bibinfo{year}{2013}), \eprint{1307.6116}.

\bibitem[{\citenamefont{Adams et~al.}(2015)\citenamefont{Adams, Ahmad, Albert,
  Allard, Anchordoqui, Andreev, Anzalone, Arai, Asano, Ave~Pernas
  et~al.}}]{jets}
\bibinfo{author}{\bibfnamefont{J.~H.} \bibnamefont{Adams}},
  \bibinfo{author}{\bibfnamefont{S.}~\bibnamefont{Ahmad}},
  \bibinfo{author}{\bibfnamefont{J.~N.} \bibnamefont{Albert}},
  \bibinfo{author}{\bibfnamefont{D.}~\bibnamefont{Allard}},
  \bibinfo{author}{\bibfnamefont{L.}~\bibnamefont{Anchordoqui}},
  \bibinfo{author}{\bibfnamefont{V.}~\bibnamefont{Andreev}},
  \bibinfo{author}{\bibfnamefont{A.}~\bibnamefont{Anzalone}},
  \bibinfo{author}{\bibfnamefont{Y.}~\bibnamefont{Arai}},
  \bibinfo{author}{\bibfnamefont{K.}~\bibnamefont{Asano}},
  \bibinfo{author}{\bibfnamefont{M.}~\bibnamefont{Ave~Pernas}},
  \bibnamefont{et~al.}, \bibinfo{journal}{Experimental Astronomy}
  \textbf{\bibinfo{volume}{40}}, \bibinfo{pages}{239} (\bibinfo{year}{2015}),
  ISSN \bibinfo{issn}{1572-9508},
  \urlprefix\url{http://dx.doi.org/10.1007/s10686-014-9431-0}.

\bibitem[{\citenamefont{{Adams} et~al.}(2015)\citenamefont{{Adams}, {Ahmad},
  {Albert}, {Allard}, {Anchordoqui}, {Andreev}, {Anzalone}, {Arai}, {Asano},
  {Ave Pernas} et~al.}}]{jemeuso_da}
\bibinfo{author}{\bibfnamefont{J.~H.} \bibnamefont{{Adams}}},
  \bibinfo{author}{\bibfnamefont{S.}~\bibnamefont{{Ahmad}}},
  \bibinfo{author}{\bibfnamefont{J.-N.} \bibnamefont{{Albert}}},
  \bibinfo{author}{\bibfnamefont{D.}~\bibnamefont{{Allard}}},
  \bibinfo{author}{\bibfnamefont{L.}~\bibnamefont{{Anchordoqui}}},
  \bibinfo{author}{\bibfnamefont{V.}~\bibnamefont{{Andreev}}},
  \bibinfo{author}{\bibfnamefont{A.}~\bibnamefont{{Anzalone}}},
  \bibinfo{author}{\bibfnamefont{Y.}~\bibnamefont{{Arai}}},
  \bibinfo{author}{\bibfnamefont{K.}~\bibnamefont{{Asano}}},
  \bibinfo{author}{\bibfnamefont{M.}~\bibnamefont{{Ave Pernas}}},
  \bibnamefont{et~al.}, \bibinfo{journal}{Experimental Astronomy}
  \textbf{\bibinfo{volume}{40}}, \bibinfo{pages}{117} (\bibinfo{year}{2015}).

\bibitem[{\citenamefont{{Neronov} et~al.}(2016)\citenamefont{{Neronov},
  {Semikoz}, {Vovk}, and {Mirzoyan}}}]{muon_showers}
\bibinfo{author}{\bibfnamefont{A.}~\bibnamefont{{Neronov}}},
  \bibinfo{author}{\bibfnamefont{D.~V.} \bibnamefont{{Semikoz}}},
  \bibinfo{author}{\bibfnamefont{I.}~\bibnamefont{{Vovk}}}, \bibnamefont{and}
  \bibinfo{author}{\bibfnamefont{R.}~\bibnamefont{{Mirzoyan}}},
  \bibinfo{journal}{ArXiv e-prints}  (\bibinfo{year}{2016}),
  \eprint{1610.01794}.

\bibitem[{\citenamefont{{Hillman} et~al.}(2003)\citenamefont{{Hillman},
  {Takahashi}, {Zuccaro}, {Lamb}, {Pitalo}, {Lopado}, {Keys}, and {EUSO
  Collaboration}}}]{euso_optics}
\bibinfo{author}{\bibfnamefont{L.~W.} \bibnamefont{{Hillman}}},
  \bibinfo{author}{\bibfnamefont{Y.}~\bibnamefont{{Takahashi}}},
  \bibinfo{author}{\bibfnamefont{A.}~\bibnamefont{{Zuccaro}}},
  \bibinfo{author}{\bibfnamefont{D.}~\bibnamefont{{Lamb}}},
  \bibinfo{author}{\bibfnamefont{K.}~\bibnamefont{{Pitalo}}},
  \bibinfo{author}{\bibfnamefont{A.}~\bibnamefont{{Lopado}}},
  \bibinfo{author}{\bibfnamefont{A.}~\bibnamefont{{Keys}}}, \bibnamefont{and}
  \bibinfo{author}{\bibnamefont{{EUSO Collaboration}}},
  \bibinfo{journal}{International Cosmic Ray Conference}
  \textbf{\bibinfo{volume}{2}}, \bibinfo{pages}{935} (\bibinfo{year}{2003}).

\bibitem[{\citenamefont{Unger et~al.}(2015)\citenamefont{Unger, Farrar, and
  Anchordoqui}}]{Unger:2015laa}
\bibinfo{author}{\bibfnamefont{M.}~\bibnamefont{Unger}},
  \bibinfo{author}{\bibfnamefont{G.~R.} \bibnamefont{Farrar}},
  \bibnamefont{and} \bibinfo{author}{\bibfnamefont{L.~A.}
  \bibnamefont{Anchordoqui}}, \bibinfo{journal}{Phys. Rev.}
  \textbf{\bibinfo{volume}{D92}}, \bibinfo{pages}{123001}
  (\bibinfo{year}{2015}), \eprint{1505.02153}.

\bibitem[{spb(2016)}]{spb}
 (\bibinfo{year}{2016}),
  \urlprefix\url{http://sites.wff.nasa.gov/code820/spb.html}.

\end{thebibliography}
%

\end{document}